\documentclass[12pt]{article}

\newcommand{\examplename}{Example}
\newcommand{\exampleref}[1]{\hyperref[#1]{\examplename~\ref*{#1}}}

\usepackage{epstopdf,amsfonts,amsmath,amssymb}
\usepackage{graphicx}
\DeclareGraphicsExtensions{.eps}
\usepackage[textsize=footnotesize]{todonotes}
\usepackage[pdftex]{hyperref}
\hypersetup{colorlinks,urlcolor=magenta,citecolor=red,linkcolor=blue,filecolor=black}

\newcommand\encadremath[1]{\vbox{\hrule\hbox{\vrule\kern8pt
\vbox{\kern8pt \hbox{$\displaystyle #1$}\kern8pt}
\kern8pt\vrule}\hrule}}
\def\enca#1{\vbox{\hrule\hbox{
\vrule\kern8pt\vbox{\kern8pt \hbox{$\displaystyle #1$}
\kern8pt} \kern8pt\vrule}\hrule}}

\usepackage[style=alphabetic,maxalphanames=4,giveninits,sorting=nyt]{biblatex}
\renewbibmacro{in:}{}
\usepackage{parskip} 
\usepackage{tikz}
\usetikzlibrary{knots}
\usetikzlibrary{calc}
\usetikzlibrary{decorations.pathreplacing} 
\usetikzlibrary{decorations.markings}
\usetikzlibrary{arrows}
\tikzset{->-/.style={decoration={
  markings,
  mark=at position .5 with {\arrow{>}}},postaction={decorate}}}

\newcommand\framefig[1]{
\begin{figure}[bth]
\hrule\hbox{\vrule\kern8pt
\vbox{\kern8pt \vbox{
\begin{center}
{#1}
\end{center}
}\kern8pt}
\kern8pt\vrule}\hrule
\end{figure}
}

\newcommand\figureframex[3]{
\begin{figure}[bth]
\hrule\hbox{\vrule\kern8pt
\vbox{\kern8pt \vbox{
\begin{center}
{\mbox{\epsfxsize=#1.truecm\epsfbox{#2}}}
\end{center}
\caption{#3}
}\kern8pt}
\kern8pt\vrule}\hrule
\end{figure}
}
\newcommand\figureframey[3]{
\begin{figure}[bth]
\hrule\hbox{\vrule\kern8pt
\vbox{\kern8pt \vbox{
\begin{center}
{\mbox{\epsfysize=#1.truecm\epsfbox{#2}}}
\end{center}
\caption{#3}
}\kern8pt}
\kern8pt\vrule}\hrule
\end{figure}
}

\makeatletter
\@addtoreset{equation}{section}
\makeatother
\newtheorem{theorem}{Theorem}[section]
\newtheorem{conjecture}{Conjecture}[section]
\newtheorem{remark}{Remark}[section]
\newtheorem{proposition}{Proposition}[section]
\newtheorem{lemma}{Lemma}[section]
\newtheorem{corollary}{Corollary}[section]
\newtheorem{definition}{Definition}[section]
\def\br{\begin{remark}\rm\small}
\def\er{\end{remark}}
\def\bt{\begin{theorem}}
\def\et{\end{theorem}}
\def\bd{\begin{definition}}
\def\ed{\end{definition}}
\def\bp{\begin{proposition}}
\def\ep{\end{proposition}}
\def\bl{\begin{lemma}}
\def\el{\end{lemma}}
\def\bc{\begin{corollary}}
\def\ec{\end{corollary}}
\def\beaq{\begin{eqnarray}}
\def\eeaq{\end{eqnarray}}
\def\bex{\begin{example}}
\def\eex{\end{example}}
\newtheorem{example}{Example}[section]
\newcommand{\proof}[1]{{\noindent \bf Proof:}\par
{#1} $\square$}

\newcommand{\be}{\begin{equation}}
\newcommand{\ee}{\end{equation}}
\newcommand{\beq}{\begin{equation}}
\newcommand{\eeq}{\end{equation}}
\newcommand{\bea}{\begin{eqnarray}}
\newcommand{\eea}{\end{eqnarray}}

\newcommand{\Tr}{{\operatorname {Tr}}}

\newcommand{\idx}{{\operatorname {index}}}

\newcommand{\ii}{{\rm i}\,}

\newcommand{\CC}{{\mathbb C}}
\newcommand{\RR}{{\mathbb R}}
\newcommand{\ZZ}{{\mathbb Z}}

\newcommand{\Tau}{{\cal T}}

\newcommand{\spcurve}{{\cal S}}
\newcommand{\curve}{{\Sigma}}

\newcommand{\curverond}{\underline{\curve}}
\newcommand{\genus}{{\mathfrak g}}

\newcommand{\acycle}{{\cal A}}
\newcommand{\bcycle}{{\cal B}}

\newcommand{\x}{{\rm x}}
\newcommand{\y}{{\rm y}}

\newcommand{\D}{\mathcal{D}}

\newcommand{\td}{\tilde}

\newcommand{\Res}{\mathop{\,\rm Res\,}}
\usepackage[pdftex]{hyperref}
\hypersetup{colorlinks,urlcolor=magenta,citecolor=red,linkcolor=blue,filecolor=black}

\begin{document}

\sloppy

\pagestyle{empty}
\addtolength{\baselineskip}{0.20\baselineskip}
\begin{center}
\vspace{26pt}
{\large \bf {Hirota, Fay and Geometry}}
\vspace{26pt}

{\sl B.\ Eynard}${}^{1,2}$\hspace*{0.05cm},
{\sl S.\ Oukassi}${}^{1}$\hspace*{0.05cm}

\vspace{6pt}
${}^{1}$Université Paris-Saclay, CNRS, CEA, Institut de physique théorique, 91191, Gif-sur-Yvette, France. \\
${}^{2}$CRM, Centre de recherches math\'ematiques  de Montr\'eal,\\
Universit\'e de Montr\'eal, QC, Canada.
\end{center}
\vspace{20pt}
\begin{center}
{\bf Abstract}
\end{center}
This is a review of the relationship between Fay identities and Hirota equations in integrable systems, reformulated in a geometric language compatible with recent Topological Recursion formalism.
We write Hirota equations as trans-series, and Fay identities as spinor functional relations. We also recall several constructions of how some solutions to Fay/Hirota equations can be built from Riemann surface geometry.

%

\vspace{0.5cm}

\vspace{26pt}
\pagestyle{plain}
\setcounter{page}{1}



\section{Introduction: Tau functions and Hirota}
\label{sec:Intro}

Tau functions are central to the theory of integrable systems, we shall refer to \cite{BBT,harnad_balogh_2021,Arthamonov_2023} for further references on integrable systems and Tau functions.

Tau functions are usually presented as functions of an infinite set of variables and are solutions of an infinite set of non-linear partial differential equations.

The most famous are ``KdV", ``KP" and ``Toda" Tau functions.

The set of equations that they satisfy is called ``Hirota equations", and can be rephrased as Fay identities.

In the form of Fay identities, the Hirota equations become equations for germs of functions of a finite number of complex variables.

It is thus natural to look for solutions of Hirota/Fay equations where the complex variables are actually local coordinates on a Riemann surface, and thus associate Tau functions to  Riemann surfaces.

We recall here some Riemannian geometric solutions of Hirota/Fay identities.

\medskip
\textbf{Outline:}

\autoref{sec:Intro} is a brief introduction to Tau functions, Sato vectors and Hirota equations in the formalism of Hirota, Sato, Segal-Wilson, and others.

\autoref{sec:trans-series}: in the usual Hirota equations, Tau functions are formal series and are multiplied by exponentials of negative powers, and Hirota equations involve a residue i.e. an integral with a differential form $d\xi$. It is thus natural to reformulate Tau functions including exponentials and differentials,  as spinor trans-series. 

\autoref{sec:Fayid}: we recall how Hirota equations lead to Fay identities. In the sense of formal series this would be equivalent, but Fay identities can be extended beyond formal series.

\autoref{sec:TauRS}: Fay identities are equations relating  germs of analytic functions, and it is thus natural to view them as equations of analytic functions on a Riemann surface.
In other words one can look for solutions of Fay identities in the world of Riemann surfaces. \autoref{sec:TauRS} is a brief summary of the geometry of Riemann surfaces.

\autoref{sec:Tau1}: we show one classical solution of Fay identities on Riemann surfaces, using Theta functions, this is the famous ``reconstruction formula" for integrable systems.

\autoref{sec:Tau2}: we show another solution of Fay identities on a moduli space of spectral curves (Riemann surface with an immersion into a cotangent space), as an asymptotic formula using ``Topological Recursion".

\autoref{sec:TauMm}: we mention other solutions, and particularly the ``matrix integrals".

\autoref{sec:conc}: we gather a number of concluding remarks.

\subsection{Tau functions as graded formal series}

A now standard formulation of Tau functions, was introduced by Sato, Miwa, Jimbo and the Japanese school \cite{JM83,Sato2,Sato1, JMI,JMII, JMIII},
and in another point of view by Segal and Wilson \cite{SW85}.
Here we stay closer to the formulation of Sato, which we present here as a {\em {formal}} Tau function that depends on an infinite set of graded coordinates called ``times"
\beq
\mathbf t = (t_1,t_2,t_3,\dots)
\qquad , \quad
\deg t_k = k.
\eeq
In the formulation of Sato and Segal-Wilson, the times are coordinates on an Abelian group $\Gamma$ of transformations acting on an infinite dimensional Grassmannian, but we shall not need these  notions here.

The Tau function 
(we denote it $\hat\Tau$ to distinguish from the slightly renormalized version $\Tau$ that we shall consider below) 
\beq
\hat\Tau(\mathbf t) \in    \CC[[\Gamma]] = \CC[[t_1,t_2,t_3,\dots]] ,
\eeq
is a formal graded series, whose truncation to any degree is a polynomial in the times, and which obeys an infinite  set of partial differential or functional equations that we shall review below.

\bd[Sato's vector]
Let $\xi\in \CC$, define the following vector (called after Sato):
\beq
[\xi] := (\xi,\xi^{2},\xi^{3},\xi^4,\dots) \in \Gamma.
\eeq
In the filtration, $\xi$ has degree $1$.
\ed

\br
Here we slightly change the usual Sato's normalization of times, by multiplying by $k$. Our times are  
\beq
t_k = k \times \text{Sato's time }t_k.
\eeq 
This multiplication by $k$ is useful for the comparison with TR (Topological Recursion) formalism, and is necessary to match the times with the canonical Darboux coordinates in the space of generalized homology cycles \cite{EynCycles}.
\er

\bd[Divisors]
A divisor $D$ is a set of distinct points $z_1,\dots,z_\ell$ of the complex plane $z_i\in\CC$, with some respective weights $\alpha_1,\dots,\alpha_\ell$, where $\alpha_i\in \CC$. Weights are also called ``charge". 
It is often denoted (this is merely a notation) as a weighted sum of points:
\beq
D :=  \sum_{i=1}^\ell \alpha_i . z_i \ .
\eeq
The sum of weigths, the total charge, is called the degree of the divisor:
\beq
\deg D := \sum_{i=1}^\ell \alpha_i.
\eeq
The support of the divisor is the set of points with non-zero weight:
\beq
\operatorname{supp}D := \{z_i \ | \ \alpha_i\neq 0\}.
\eeq
The set of divisors is a module by addition and scalar multiplication of the weights.

A divisor $D$ is said \textbf{neutral} if $\deg D=0$. 
A divisor is said \textbf{integer} if weights are integers, and is \textbf{unitary} if weights are $\pm 1$.
A divisor is said \textbf{positive} or \textbf{bosonic} (resp. \textbf{negative} or \textbf{fermionic}) if all weights are positive (resp. negative).
A divisor is said \textbf{supsersymmetric} if it is of degree $0$ and all weights are $\pm 1$. A supersymmetric divisor must have an even number of points, half of them have weight $+1$ (bosons) and half of them have weight $-1$ (fermions).
\ed

\bd[Sato's vector for divisors]
We associate a Sato vector to a divisor $D=\sum_i \alpha_i.z_i $, equal to the weighted linear combination of Sato vectors of its points:
\beq
[D] = \sum_{i=1}^\ell \alpha_i [z_i] = (D_1,D_2,D_3,\dots)
\in \Gamma
\quad \text{where}\quad
D_k = \sum_{i=1}^\ell \alpha_i z_i^{k}.
\eeq
It is convenient to think of $D$ as a diagonal matrix with eigenvalues $z_i$ and multiplicities $\alpha_i$ ($\alpha_i<0$ means ``super-matrix" with ``fermionic" eigenvalues).
With the diagonal matrix notation we have:
\beq
D_k = \Tr D^{k}.
\eeq

\ed

\br[\textbf{Screening:} Divisors of degree 0]
Remark that if $z=0$, then the Sato vector $[z] = (0,0,0,\dots)$ is the null vector in $\Gamma$.
Therefore if a divisor $D=\sum_{i=1}^\ell \alpha_i [z_i]$ has a non-zero degree, $\deg D\neq 0$, we can always suplement it by appending to it the point 
$$z_0=0, \qquad \alpha_0=-\deg D.$$
We get a new divisor whose degree is 0
\beq
\td D=D+\alpha_0.0, \quad \deg \td D=0
\eeq
and such that
\beq
\forall \ k>0 \ \qquad \td D_k=D_k=\Tr \td D^k =\Tr D^k,
\eeq
i.e. as a vector in $\Gamma$
\beq
[\td D] = (D_1,D_2,D_3,\dots) = [D].
\eeq
In other words it is not restrictive to require that a divisor has degree 0, it can be completed by the point $0$ with the complementary weight. This is  called ``screening charge".

\er

\subsection{Hirota Equation}

Hirota equation is a set of non-linear PDEs satisfied by $\hat\Tau$, and gathered into a bilinear relation that generates all of them:

\bd[Hirota Equation]
$\hat\Tau(\vec t)$ is said to satisfy Hirota equation iff for all 
$\ell$ a positive integer, and all $\mu=(\mu_1,\mu_2,\dots,\mu_\ell)\in \mathbb N^\ell$ a $\ell$-upple, we have
\beq
\Res_{\xi\to 0} [u_1^{\mu_1} u_2^{\mu_2}\dots u_\ell^{\mu_\ell}] \left(
\hat\Tau(\vec t + \vec u+ [\xi]) \hat\Tau(\vec t-\vec u - [\xi]) 
e^{- 2\sum_{k=1}^{\ell} \frac{u_k}{k}\xi^{-k} }
\right) \frac{d\xi}{\xi^2}
= 0.
\eeq
The notation $[u^\mu](.)= [u_1^{\mu_1} u_2^{\mu_2}\dots u_\ell^{\mu_\ell}] \left(. \right)$ means that we first expand the parenthesis as a power series of the $u_k$s and extract the coefficient of the monomial $u_1^{\mu_1} u_2^{\mu_2}\dots u_\ell^{\mu_\ell}$, and then we compute the residue at $\xi\to 0$.

\ed
This is well defined, because the exponential is equal to its  Taylor series in the $u_k$s (absolutely convergent moreover), and since $\hat\Tau$ is a graded formal series, for a given $\ell$-upple $\mu$, there is at most a finite number of terms contributing to the monomial $[u^\mu]$. The expression is then a polynomial of $\xi^{-1}$, times a formal power series of $\xi$, therefore its residue is well defined (the coefficient of $\xi^{-1}$).

\medskip

In fact \textbf{Hirota} introduced it, because he had realized that this equation was common to almost all integrable systems known at that time:

\bt[Hirota]
The Tau functions of the KdV, KP, Toda hierarchies satisfy the Hirota equation.

\et
For the proof we refer to the litterature   
\cite{HR71,Hirota, BBT, harnad_balogh_2021,Arthamonov_2023}.

\subsubsection{Turning Hirota to PDEs}

Let us review \cite{BBT,harnad_balogh_2021,Arthamonov_2023} how Hirota equation yields an infinite set of non-linear PDEs.

We have, in the sense of formal series in $\CC[[\xi,u_1,u_2,\dots]]$, the Taylor formula:
\beq
\hat\Tau(\vec t + \vec u+ [\xi])
= e^{\sum_{k=1}^\infty (u_k+\xi^{k})\frac{\partial}{\partial t_k} }\hat \Tau(\vec t).
\eeq
Since Hirota equation is bilinear and involves two Tau functions, some derivatives will act onto the 1st Tau function, and some onto the 2nd. 
We shall write one Tau function at the left, the other at the right, and define two operators $\overset\leftarrow{\frac{\partial}{\partial t_k}}$ and $\overset\rightarrow{\frac{\partial}{\partial t_k}}$.
We also introduce the differential bi-operator :
\beq
\D_k := \overset\leftarrow{\frac{\partial}{\partial t_k}} - \overset\rightarrow{\frac{\partial}{\partial t_k}}.
\eeq
This means that in $\CC[[\xi,u_1,u_2,\dots]]$
\bea
\hat\Tau(\vec t + \vec u+ [\xi])
\hat\Tau(\vec t - \vec u- [\xi])
&=& \hat \Tau(\vec t) e^{\sum_k (u_k+\xi^{k})\overset\leftarrow{\frac{\partial}{\partial t_k}} }
 e^{-\sum_k (u_k+\xi^{k})\overset\rightarrow{\frac{\partial}{\partial t_k}} }\hat \Tau(\vec t) \cr
&=& \hat \Tau(\vec t) e^{\sum_k (u_k+\xi^{k}) 
D_k }\hat \Tau(\vec t) .
\eea
Hirota equations can then be written as follows
\bt[Hirota as PDEs]
For all $\mu=(\mu_1,\dots,\mu_\ell)$
\beq
\hat\Tau(\vec t) \ \D_\mu \ \hat\Tau(\vec t) 
= 0
\eeq
where $\D_\mu$ is a differential bi-operator defined as
\beq
\D_\mu =  \left( \Res_{\xi\to 0} \frac{d\xi}{\xi^2} \  e^{\sum_k \xi^{k}D_k} \ 
\prod_{j=1}^\ell \frac{\left(\D_j-\frac{2}{j}\xi^{-j}\right)^{\mu_j}}{\mu_j!}\right).
\eeq
$\D_\mu$ is a polynomial of $\D_1,\D_2,\D_3,\dots$ of graded homogeneous degree $1+\sum_{i=1}^\ell i\mu_i$.
More explicitely:
\beq
\D_\mu
= \sum_{\sum_i i m_i + \sum_j j\alpha_j = 1+\sum_j j \mu_j} \ \ 
\prod_{j=1}^\ell \left(\frac{-2}{j}\right)^{\mu_j-\alpha_j} \ \frac{\D_j^{\alpha_j}}{\alpha_j! (\mu_j-\alpha_j)!} \ \prod_{i\geq 1} \frac{\D_i^{m_i}}{m_i!}
\eeq

\et

\br
Remark that $\hat\Tau(\vec t) \ \D_\mu \ \hat\Tau(\vec t)$ is symmetric in the exchange of the $\Tau$ at the left and at the right.
This implies that $\hat\Tau(\vec t) \ \D_\mu \ \hat\Tau(\vec t) 
$ is unchanged if we change each $\D_k\to -\D_k$.
In other words, we may keep in $\D_\mu$ only the terms that are even in the change $\D_k\to -\D_k$.
Any odd polynomial of  $\D_1,\D_2,\D_3,\dots$ will trivially gives 0 in the equation $\hat\Tau(\vec t) \ \D_\mu \ \hat\Tau(\vec t)$.

\er

\bex

\beq
\D_{(0,0,0,\dots)}
= \D_1, \quad
\D_{(1,0,0,\dots)}
= -2\D_2,\quad
\D_{(0,1,0,\dots)}
= -\frac16 \D_1^3-\D_3
\eeq
\beq
\D_{(2,0,0,\dots)}
= -\frac16 \D_1^3+2\D_3,\quad
\D_{(0,0,1,0,\dots)}
=-\frac{\D_{1}^{4}}{36} - \frac{\D_{1}^{2} \D_{2}}{3} + \frac{\D_{1} \D_{3}}{3} - \frac{\D_{2}^{2}}{3} - \frac{2 \D_{4}}{3}
\eeq
\beq
\D_{(2,1,0,\dots)}
= -\frac{\D_{1}^{5}}{60} + \frac{\D_{1}^{2} \D_{3}}{2} - 2 \D_{5}.
\eeq

For example, $\hat\Tau \D_{(0,0,1,0,\dots)}\hat\Tau = 0$ implies the following differential equation
\bea
0 &=& 
\hat\Tau \ \partial_{t_2}^2 \hat\Tau - \partial_{t_2}\hat\Tau \ \partial_{t_2}\hat\Tau \cr
&& +\frac{1}{12}\Big(\hat\Tau \ \partial{t_1}^4 \hat\Tau - 4 \partial_{t_1}\hat\Tau \ \partial{t_1}^3 \hat\Tau + 3 \partial_{t_1}^2\hat\Tau \ \partial{t_1}^2 \hat\Tau \Big) \cr
&& - \hat\Tau \ \partial{t_1}\partial_{t_3} \hat\Tau +  \partial{t_1} \hat\Tau \ \partial_{t_3} \hat\Tau
\eea
If we write $\hat\Tau = e^{\hat F}$ we get the KP equation
\bea
0 &=&
\partial_{t_2}^2 \hat F   +\frac{1}{12}\Big(\partial_{t_1}^4 \hat F+6 (\partial_{t_1}\hat F)^2 \Big)  - \partial_{t_3}\partial_{t_1} \hat F.
\eea
\eex

\subsection{Insertion operator}

\bd[Insertion operator (perturbative)]We define
\label{def:InsertionOp}
\beq
\label{eqdef:InsertionOp}
\Delta_\xi :=   d\xi\sum_{k=1}^\infty k \xi^{k-1} \frac{\partial}{\partial t_k} = d\left( \sum_{k=1}^\infty \xi^{k} \frac{\partial}{\partial t_k} \right).
\eeq
$\Delta$ is a 1-form valued differential operator.
As defined here, it acts on the space $\CC[[\Gamma]]$ of graded formal series of the times, and results in a formal series of $\xi$, times $d\xi$:
\beq
\Delta : \CC[[t_1,t_2,\dots]]  \to \CC[[t_1,t_2,\dots]][[\xi]] d\xi.
\eeq
\ed

\bl[Insertion operator and shifted times]
\label{lem:DeltavsShifts}
For any $f(\vec t)\in \CC[[\Gamma]] $, we have (in the sense of formal graded series)
\beq\label{eq:shiftexpDelta}
f(\vec t+\alpha[\xi])
 = e^{\alpha\int_0^\xi \Delta} f(\vec t),
\eeq
\beq\label{eq:Deltalimitshift}
\Delta_\xi f(\vec t)  = \lim_{\alpha\to 0}\left(\frac{1}{\alpha}d\xi \ \frac{d}{d\xi}f(\vec t+\alpha [\xi])\right).
\eeq

\beq\label{eq:shiftexpDeltadeg0}
f(\vec t+\alpha[\xi]-\alpha[\td\xi])
 = e^{\alpha\int_{\td\xi}^\xi \Delta} f(\vec t),
\eeq
\beq\label{eq:Deltalimitshiftdeg0}
\Delta_\xi f(\vec t)  = \lim_{\alpha\to 0}\left(\frac{1}{\alpha}d\xi \ \frac{d}{d\xi}f(\vec t+\alpha [\xi]-\alpha[\td\xi])\right).
\eeq

\el

\proof{
From \autoref{def:InsertionOp} we can write:
\bea
e^{\alpha\int_{\td\xi}^\xi \Delta} f(\vec t)
&=& e^{\sum_{k=1}^{\infty}\alpha \left(\xi^k-\td\xi^k\right) \frac{\partial}{\partial t_k} }  \  f(t_1,t_2,\dots) \cr
&=&  f(\vec t + \alpha[\xi]-\alpha[\td\xi]).
\eea
The exponentiel term, viewed as an element of $\mathbb{C}[[\xi,\td\xi]]$, can be written as:
\beq
e^{\sum_{k=1}^{\infty}\alpha \left(\xi^k-\td\xi^k\right) \frac{\partial}{\partial t_k} }f(\vec t)=\sum_{l=0}^{\infty} \frac{\alpha^l}{l!}\left(\sum_{k=1}^{\infty} (\xi^k-\td\xi^k) \frac{\partial}{\partial t_k} \right)^l f(\vec t).
\eeq
\beq
\frac{d}{d\xi}f(\vec t+\alpha [\xi]-\alpha[\td\xi])=\sum_{l=1}^{\infty} \frac{\alpha^l}{(l-1)!}\left(\sum_{k=1}^{\infty} (\xi^k-\td\xi^k) \frac{\partial}{\partial t_k} \right)^{l-1} \left(\sum_{k=1}^{\infty} k \xi^{k-1} \frac{\partial}{\partial t_k}\right)f(\vec t).
\eeq
Hence, we have
\beq
\lim_{\alpha\to 0}\left(\frac{1}{\alpha}d\xi \ \frac{d}{d\xi}f(\vec t+\alpha [\xi]-\alpha[\td\xi])\right)=\sum_{k=1}^{\infty} k \xi^{k-1} \frac{\partial}{\partial t_k}f(\vec t)=\Delta_\xi f(\vec t).
\eeq
Moreover, eq \eqref{eq:Deltalimitshift} is obtained from eq \eqref{eq:Deltalimitshiftdeg0} by taking the limit $\td\xi \to 0$, as well as eq \eqref{eq:shiftexpDelta}.
}
\br Eq~\eqref{eqdef:InsertionOp} is the Taylor series expansion of eq~\eqref{eq:Deltalimitshiftdeg0} or \eqref{eq:Deltalimitshift}. 
From now on, we shall use eq~\eqref{eq:Deltalimitshiftdeg0} as the definition of the insertion operator $\Delta$ instead of eq~\eqref{eqdef:InsertionOp}, because it is now defined not only as a germ but as a genuine analytic function of $\xi$, and it can act on functions of complex variables rather than formal series of times.
\er

\bd[Potential]
Let
\beq
V_{\vec t}(\xi) :=\sum_{k=1}^\infty \frac{t_k}{k} \xi^{-k}  \in \CC[[\xi^{-1}]].
\eeq

We have
\beq
dV_{\vec t}(\xi) = -\sum_{k=1}^\infty t_k \xi^{-k-1}d\xi .
\eeq
\beq
\label{eq:deltashiftpot}
\Delta_{\xi_2}dV_{\vec t}(\xi_1) = - \frac{d\xi_1 \otimes d\xi_2}{(\xi_1-\xi_2)^2}.
\eeq
\ed
\br
 If times grow like $|t_k|=O(R^{k})$ at large $k$ for some $R>0$, then $V_{\vec t}(\xi)$ is analytic in a disc around $\infty$ of radius $|\xi|\geq R$.
\er

\bd[Correlators]
\label{def:Correlators}

The following quantities are called the ``correlators":
\beq
\hat W_0(\vec t) := \ln\hat\Tau(\vec t)
\eeq
\beq
W_1(\vec t;\xi) := \Delta_\xi \ln\hat\Tau(\vec t) \  +\sum_{k\geq 1} t_k \xi^{-k-1}d\xi 
= \Delta_\xi \ln\hat\Tau(\vec t) \  -dV_{\vec t}(\xi)
\eeq
\beq
W_2(\vec t;\xi_1,\xi_2) := \Delta_{\xi_1}\otimes\Delta_{\xi_2} \ln\hat\Tau(\vec t) \  + \frac{d\xi_1 \otimes d\xi_2}{(\xi_1-\xi_2)^2}.
\eeq
and for $n\geq 3$
\beq
W_n(\vec t;\xi_1,\xi_2,\dots,\xi_n) := \Delta_{\xi_1}\otimes\Delta_{\xi_2}\otimes \dots \otimes \Delta_{\xi_n} \ln\hat\Tau(\vec t) .
\eeq

\ed

\bl
\label{lemma:derivTauWn}
For $k>0$
\beq
t_k =  \Res_{\xi\to 0} \xi^{k} \ W_1(\vec t;\xi),
\eeq
\beq
\frac{\partial}{\partial t_k}\ln\hat\Tau(\vec t) = \frac{1}{k} \Res_{\xi\to 0} \xi^{-k} \ W_1(\vec t;\xi)
\eeq
For $k_1,k_2,\dots,k_n>0$
\beq
\frac{\partial}{\partial t_{k_1}}\dots \frac{\partial}{\partial t_{k_n}}\ln\hat\Tau(\vec t) = \frac{1}{k_1\dots k_n} \Res_{\xi_1\to 0} \dots \Res_{\xi_n\to 0} \xi_1^{-k_1}\dots \xi_n^{-k_n} \ W_n(\vec t;\xi_1,\dots,\xi_n).
\eeq
\el
\proof{
Direct consequences of the definitions.
}
\section{Reformulation as spinor trans-series}
\label{sec:trans-series}
\subsection{Motivation}

Let us make some observations:

\begin{itemize}

\item A key ingredient in Hirota equation, is that a shift $\vec t\mapsto \vec t+\vec u +[\xi]$ is understood as a formal series in $u_k$s and in $\xi$, computed by the Taylor series formula.

\item in Hirota equation, the Sato-shifted Tau functions always appear  multiplied by an exponential, which is not a power series of $\xi$, instead it contains exponential of negative powers of $\xi$ (and which are useful to produce residues). 
Somehow, shifting is accompanied by an exponential term. 

Formal series multiplied by exponentials of negative powers are not formal series, they are called ``\textbf{trans-series}" (see \cite{Ecalle1,Ecalle2,Ecalle3, MSb16,sauzin2014introduction}).

\item  The insertion operator $\Delta$ is formally an operator of degree $0$ in the filtration (indeed $\partial /\partial t_k$ has degree $-k$ while $\xi^k$ has degree $k$), but is made of infinitely many terms. 
This means that associativity and commutativity of using $e^{\alpha\int\Delta}$ is not guaranteed. In some sense such operators need to be defined outside of the ring $\CC[[\Gamma]]$ of graded formal series, they need to take into account exponential prefactors.

\item Another observation is that Residues are defined for 1-forms, and they are invariant under change of variables only if one includes the Jacobian coming from the $d\xi$. In other words, the nature of the product of Tau functions inside the residue, should be a 1-form, which means that each factor $\hat\Tau$ transforms as a power of $d\xi$, such that the product of the two factors transforms as $d\xi$. We shall redefine $\hat\Tau$ to include a power of $d\xi$, this will make $\hat\Tau$ a spinor (a differential with fractional power of $d\xi$), typically each $\hat\Tau$ will carry a half-integer power $\sqrt{d\xi}$.

\item
In actual realizations of integrable systems, for example as integrals (example matrix integrals in random matrix theory, see \autoref{sec:matrices} below), Tau functions have asymptotic expansions that often are not  convergent series, they are usually factorially divergent series, and they usually have an exponential prefactor, which is not a series in $\xi$. 
The good formalism to describe asymptotic expansions of exponential integrals or random matrix integrals, is again the notion of trans--series rather than series.
Moreover, the theory of resurgence \cite{Ecalle1,Ecalle2,Ecalle3, MSb16,sauzin2014introduction}, claims that whenever we have a divergent series, it should always be considered as trans--series.

\end{itemize}

\smallskip

All this suggest that the good framework for Tau functions should not be ``series", but rather ``trans-series", and moreover spinors (with a $\sqrt{d\xi}$).

Trans-series are linear combinations of series multiplied by some exponential. They have a rather different behavior than series, and in particular the Taylor formula and other operations can't be applied naively.
This shows that Hirota equations should better be reformulated in \textbf{spinor trans-series} rather than series.

\subsection{Tau function as trans-series}

\bd[Tau function]
\label{def:Tauts}

We define a  Tau function $\Tau$ by the following functional properties:
we say that $\Tau$ is a ``Tau function" iff:

\begin{itemize}

\item 
\beq
\label{eq:expFtau}
\Tau := e^F \hat\Tau
\eeq
where $\hat\Tau(\vec t) \in \CC[[\Gamma]] $ is a graded formal series, and $F$ is a function on $\Gamma$ homogeneous  of degree 2:
\beq
F(\lambda \vec t) = \lambda^2 F(\vec t),
\eeq
(but $F\notin \CC[[\Gamma]]$),

\item for every divisor $D=\sum_{i=1}^\ell \alpha_i. z_i$ of degree $\deg D=0$:
\beq
\label{eq:tau0div}
\Tau(\vec t+[D])
=  \Tau(\vec t) e^{-\sum_{k\geq 1} t_k \sum_i \alpha_i \frac{z_i^{-k}}{k}} \prod_{i<j} E(z_i,z_j)^{\alpha_i\alpha_j} \ \frac{\hat \Tau(\vec t+[D])}{\hat \Tau(\vec t)} 
\eeq
where $E(z_i,z_j)$ is the Fay's prime form \cite{Fay}:
\beq
E(z,z') := \frac{z-z'}{\sqrt{ dz dz'}} ,
\eeq
and where $\hat\Tau(\vec t+[D])$ is a graded formal series
\beq
\hat\Tau(\vec t+[D]) \in \CC[[z_1,\dots,z_\ell,\vec t]].
\eeq

\item It satisfies the Hirota equation (see \autoref{prop:HirotaforTau} below), or its alternative Fay equations \autoref{th:HirotaDivisors} or \autoref{th:HirotaFay} below.

\end{itemize}
\ed

\bp
The definition is consistent. With this definition, the Sato shift is associative (and commutative up  to a phase):
if $D$ and $D'$ are two divisors of degree 0 we have
\beq
\Tau((\vec t+[D])+[D']) = \Tau(\vec t+([D]+[D'])). 
\eeq
For $\sigma\in \mathfrak S_{\ell} $ a permutation, we denote $\sigma(D) = \sum_{i=1}^{\ell} \alpha_{\sigma(i)}.z_{\sigma(i)}$ the permuted divisor. We have
\beq
\Tau(\vec t+[\sigma(D)]) = e^{\pi i\sum_{i<j, \ \sigma(i)>\sigma(j)}\alpha_i\alpha_j }  \ \ \Tau(\vec t+[D]) .
\eeq
If all charges $\alpha_i\in 2\mathbb Z+1$ are odd, the Tau function is fermionic, it obeys the Pauli symmetry, namely:
\beq
\Tau(\vec t+[\sigma(D)]) = (-1)^\sigma \ \Tau(\vec t+[D]) .
\eeq
\ep
\proof{
\bea
\Tau((\vec t+[D])+[D']) 
=&\Tau(\vec t+[D]) e^{-\sum_k (t_k+\sum_j \alpha_j z_j^k) \sum_i \alpha'_i \frac{{z'_i}^{-k}}{k}} \cr
& \prod_{i<j} E(z'_i,z'_j)^{\alpha'_i\alpha'_j} \ \frac{\hat \Tau(\vec t+[D]+[D'])}{\hat \Tau(\vec t+[D])} \cr
=&\Tau(\vec t+[D])
 \prod_i e^{-\sum_k t_k \alpha_i' \frac{{z'_i}^{-k}}{k}} 
\prod_{i,j} e^{-\sum_k \alpha'_i\alpha_j z_j^k\frac{{z'_i}^{-k}}{k}} \cr
& \prod_{i<j} E(z'_i,z'_j)^{\alpha'_i\alpha'_j} \ \frac{\hat \Tau(\vec t+[D]+[D'])}{\hat \Tau(\vec t+[D])} \cr
=&\Tau(\vec t+[D])
 \prod_i e^{-\sum_k t_k \alpha_i' \frac{{z'_i}^{-k}}{k}} 
\prod_{i,j} e^{\alpha'_i\alpha_j \ln(1-z_j/z'_i)} \cr
& \prod_{i<j} E(z'_i,z'_j)^{\alpha'_i\alpha'_j} \ \frac{\hat \Tau(\vec t+[D]+[D'])}{\hat \Tau(\vec t+[D])} \cr
=&\Tau(\vec t+[D]) 
 \prod_i e^{-\sum_k t_k \alpha_i' \frac{{z'_i}^{-k}}{k}} 
\prod_{i,j} (1-z_j/z'_i)^{\alpha'_i\alpha_j } \cr
& \prod_{i<j} E(z'_i,z'_j)^{\alpha'_i\alpha'_j} \ \frac{\hat \Tau(\vec t+[D]+[D'])}{\hat \Tau(\vec t+[D])} \cr
=&\Tau(\vec t+[D])\
 \prod_i z_i'^{-\alpha'_i \deg D} e^{-\sum_k \alpha_i' t_k  \frac{{z'_i}^{-k}}{k}} 
\prod_{i,j} (z'_i-z_j)^{\alpha'_i\alpha_j } \cr
& \prod_{i<j} E(z'_i,z'_j)^{\alpha'_i\alpha'_j} \ \frac{\hat \Tau(\vec t+[D]+[D'])}{\hat \Tau(\vec t+[D])} \cr
=&\Tau(\vec t)  
\prod_j e^{-\alpha_j \sum_k \frac{t_k}{k} z_j^{-k} } \  
 \prod_{i<j} E(z_i,z_j)^{\alpha_i \alpha_j}  \frac{\hat\Tau(\vec t+[D])}{\hat \Tau(\vec t)} \cr
& \prod_i e^{-\sum_k t_k  \alpha_i' \frac{{z'_i}^{-k}}{k}} 
\prod_{i,j} E(z'_i,z_j)^{\alpha'_i\alpha_j }  \cr
& \prod_i dz_i'^{\frac12 \alpha'_i \deg D}\prod_j dz_j^{\frac12 \alpha_j \deg D'}  \cr
& \prod_{i<j} E(z'_i,z'_j)^{\alpha'_i\alpha'_j} \ \frac{\hat \Tau(\vec t+[D]+[D'])}{\hat \Tau(\vec t+[D])} \cr
=&\Tau(\vec t) \prod_j e^{-\alpha_j \sum_k \frac{t_k}{k} z_j^{-k} }  \prod_i e^{-\sum_k t_k \alpha_i' \frac{{z'_i}^{-k}}{k}} \cr
&\prod_{i<j} E(z_i,z_j)^{\alpha_i \alpha_j}  \prod_{i,j} E(z'_i,z_j)^{\alpha'_i\alpha_j }  \prod_{i<j} E(z'_i,z'_j)^{\alpha'_i\alpha'_j} \cr
&  \ \frac{\hat \Tau(\vec t+[D]+[D'])}{\hat \Tau(\vec t)} \cr
=& \Tau(\vec t + ([D]+[D'])) 
\eea
which concludes the associativity.

Now let us consider a permutation $\sigma$.
We have: 
\beq
\Tau(\vec t+[\sigma(D)])=\Tau(\vec t) e^{-\sum_k t_k \sum_i \alpha_{\sigma(i)} \frac{z_{\sigma(i)}^{-k}}{k}} \prod_{i<j} E(z_{\sigma(i)},z_{\sigma(j)})^{\alpha_{\sigma(i)}\alpha_{\sigma(j)}} \ \frac{\hat \Tau(\vec t+[D])}{\hat \Tau(\vec t)},
\eeq
furthermore
\bea
\prod_{i<j} E(z_{\sigma(i)},z_{\sigma(j)})^{\alpha_{\sigma(i)}\alpha_{\sigma(j)}} &=&\prod_{i<j, \sigma(i)>\sigma(j)} E(z_{\sigma(i)},z_{\sigma(j)})^{\alpha_{\sigma(i)}\alpha_{\sigma(j)}} \cr
&& \prod_{i<j, \sigma(i)<\sigma(j)} E(z_{\sigma(i)},z_{\sigma(j)})^{\alpha_{\sigma(i)}\alpha_{\sigma(j)}}\cr
&=&\prod_{i<j, \sigma(i)>\sigma(j)} (-E(z_{\sigma(j)},z_{\sigma(i)}))^{\alpha_{\sigma(j)}\alpha_{\sigma(i)}} \cr
&& \prod_{i<j, \sigma(i)<\sigma(j)} E(z_{\sigma(i)},z_{\sigma(j)})^{\alpha_{\sigma(i)}\alpha_{\sigma(j)}}\cr
&=&e^{\pi i\sum_{i<j, \ \sigma(i)>\sigma(j)}\alpha_i\alpha_j }\cr 
&& \prod_{i\neq j, \sigma(i)<\sigma(j)} E(z_{\sigma(i)},z_{\sigma(j)})^{\alpha_{\sigma(i)}\alpha_{\sigma(j)}}\cr
&=&e^{\pi i\sum_{i<j, \ \sigma(i)>\sigma(j)}\alpha_i\alpha_j }\prod_{i<j} E(z_i,z_j)^{\alpha_i\alpha_j}.\cr
\eea
Therefore,
\beq
\Tau(\vec t+[\sigma(D)]) = e^{\pi i s_\sigma} \ \ \Tau(\vec t+[D]) 
\eeq
where
\beq
s_\sigma = \sum_{i<j, \ \sigma(i)>\sigma(j)}\alpha_i\alpha_j   .
\eeq
Notice that the parity of $s_\sigma$ is the same as the parity of only the terms with odd $\alpha_i$s.

Assume that all $\alpha_i$s are odd, i.e. $\alpha_i \equiv 1 \mod 2$.
Then we have
\beq
s_\sigma \equiv \sum_{i<j, \ \sigma(i)>\sigma(j)} 1  \mod 2,
\eeq
and we recognize the number of inversions of $\sigma$.
It is well known that the parity of the number of inversions  is the signature of $\sigma$, therefore, when all $\alpha_i$ are odd, we have
\beq
\Tau(\vec t+[\sigma(D)]) = (-1)^{s_\sigma} \ \ \Tau(\vec t+[D]). 
\eeq

}
\bp

\beq
\label{eq:F0div}
F(\vec t+D) = F(\vec t) -\sum_{k=1}^\infty \frac{1}{k} t_k \Tr D^{-k}
+ \sum_{i<j} \alpha_i\alpha_j \ln{E(z_i,z_j)}
\eeq
\beq
\Delta_\xi F = -dV_{\vec t}(\xi).
\eeq
\beq
\Delta_\xi \ln\Tau = W_1(\xi).
\eeq
\beq
\label{eq:Deltashiftcorr}
\Delta_{\xi_1}\dots \Delta_{\xi_n} \ln\Tau = W_n(\xi_1,\dots,\xi_n).
\eeq

\ep
\proof{
 Let's take a zero degree divisor, we have:
 \beq
 \frac{\Tau(\vec t+D)}{\Tau(\vec t)}=e ^{F(\vec t+ D)-F(\vec t)} \,\frac{\hat\Tau(\vec t+D)}{\hat\Tau(\vec t)}.
 \eeq

From eq \eqref{eq:tau0div} we write

\beq
F(\vec t+D)-F(\vec t)=-\sum_{k\geq 1}t_k \sum_i \alpha_i \frac{z_i^{-k}}{k}+\sum_{i<j} \alpha_i \alpha_j \ln E(z_i,z_j).
\eeq

For $k>0$, $\Tr D^{-k}=\sum_{i=1}^{l} \alpha_i z_i^{-k}$, therefore 
\beq
F(\vec t+D) = F(\vec t) -\sum_{k=1}^\infty \frac{1}{k} t_k \Tr D^{-k}
+ \sum_{i<j} \alpha_i\alpha_j \ln{E(z_i,z_j)}.
\eeq

Consider the following divisor
\beq
D=\alpha [\xi]-\alpha[\xi_1]
\eeq
where $\xi_1$ is a fixed complex number. $D$ is a zero degree divisor and therefore eq \eqref{eq:F0div} gives 

\beq
F(\vec t+ \alpha [\xi]-\alpha[\xi_1])=F(\vec t)- \sum_{k=1}^{\infty}\frac{1}{k}t_k\left( \alpha \xi^{-k}-\alpha \xi_1^{-k}\right)-\alpha^2\ln E(\xi,\xi_1).
\eeq 
Using eq \eqref{eq:Deltalimitshiftdeg0} of \autoref{lem:DeltavsShifts} we get
\beq
\Delta_{\xi} F=\sum_{k=1}^{\infty} t_k \xi^{-k-1} d\xi=-dV_{\vec t} (\xi).
\eeq

Eq \eqref{eq:tau0div} implies

\beq
\begin{split}
\Delta_\xi \ln \Tau&=\Delta_\xi \ln \hat \Tau+\Delta_\xi F\\
&=\Delta_\xi \ln \hat \Tau-dV_{\vec t} (\xi)\\
&=W_1(\vec t, \xi).
\end{split}
\eeq
 Let $\xi_1$, $\xi_2$  $\in \CC$, we have
 
\beq
\Delta_{\xi_2} \otimes \Delta_{\xi_1} \ln \Tau=\Delta_{\xi_2} \otimes \Delta_{\xi_1} \ln \hat \Tau+\Delta_{\xi_2} \otimes \Delta_{\xi_1} F.
\eeq
Using eq \eqref{eq:deltashiftpot} we write
\beq
\begin{split}
\Delta_{\xi_2} \otimes \Delta_{\xi_1} F&=\Delta_{\xi_2} (\Delta_{\xi_1} F)\\
&=\Delta_{\xi_2} (-dV_{\vec t} (\xi))\\
&= \frac{d\xi_1 \otimes d\xi_2}{(\xi_1-\xi_2)^2}.
\end{split}
\eeq
From this last equality we can see that the quantity $\Delta_{\xi_2} \otimes \Delta_{\xi_1} F$ is independant of $\vec t$, therefore for $n>2$ we get
\beq
\Delta_{\xi_1}\dots \Delta_{\xi_n} F=0,
\eeq
thus eq \eqref{eq:Deltashiftcorr} follows.
}

\br[$F$]
In some sense, if one could define some negative times $t_{-k}$, whose Sato shift would be $t_{-k}\to t_{-k}+\Tr D^{-k}$, one would have that $F=\frac12 \sum_{k=1}^\infty \frac{1}{k} t_k t_{-k}$. 
However this is ill-defined here, it can make sense only if one would introduce a pairing between $C[[t_1,t_2,\dots]]$ and $C[[t_{-1},t_{-2},\dots]]$, and this can be done case by case in various integrable systems. See the function $F_0$ in \cite{Eyn17,EynCycles}, and that it requires a choice of polarization.
So here instead, we define $F$ by its functional property eq~\eqref{eq:F0div}.
\er

\bp[Hirota]
\label{prop:HirotaforTau}
The Hirota equation can be written
\bea
\Res_{\xi\to 0} 
\Tau(\vec t + \vec u+ [\xi]) \Tau(\vec t-\vec u - [\xi]) 
= 0 \quad \text{in } \CC[[u_1,u_2,\dots]] 
\eea
or equivalently $\forall \mu=(\mu_1,\dots,\mu_\ell)$
\bea
\Res_{\xi\to 0} [u_1^{\mu_1} u_2^{\mu_2}\dots u_n^{\mu_n}] \left(
\Tau(\vec t + \vec u+ [\xi]) \Tau(\vec t-\vec u - [\xi]) 
\right) 
= 0
\eea
or equivalently
\bea
\Res_{\xi\to 0} \Res_{u_1\to 0}\dots\Res_{u_n\to 0} \frac{du_1}{u_1^{\mu_1+1}} \frac{du_2}{u_2^{\mu_2+1}}\dots \frac{du_n}{u_n^{\mu_n+1}} \left(
\Tau(\vec t + \vec u+ [\xi]) \Tau(\vec t-\vec u - [\xi]) 
\right) 
= 0.
\eea

\ep
\proof{Simple rewritting since  our definition of $\Tau$ consisted in including the exponential and the $d\xi$ in the definition of $\Tau$.}

\subsection{Hirota with Divisors}

The goal is to choose $\vec u$  as the Sato vector of a divisor $D$ of degree $\deg D=-1$ (so that $\vec u+[\xi]$ is a divisor of degree 0):
\beq
\vec u = [D] = \sum_{j=1}^\ell \alpha_j [z_j], \quad \deg D = \sum_{j=1}^\ell \alpha_j = -1,
\eeq
and replace in Hirota equations  the formal series expansion in powers of $u^\mu$ as formal series in powers of $z_j$s.

\bt[Hirota with divisors]

We have the equivalence between

\begin{itemize}

\item 
\beq\label{eq:equivH10}
0=\Res_{\xi\to 0}\frac{d\xi}{\xi^2} \ \hat \Tau(\vec t+\vec u+[\xi]) \hat \Tau(\vec t-\vec u-[\xi]) e^{-2\sum_k \frac{u_k}{k} \xi^{-k}} 
\quad \text{in } \CC[[u_1,u_2,\dots]] 
\eeq
or equivalently
\beq\label{eq:equivH1}
\forall \mu \ , 
 \quad
0=\Res_{\xi\to 0}\frac{d\xi}{\xi^2} [\vec u^\mu] \hat \Tau(\vec t+\vec u+[\xi]) \hat \Tau(\vec t-\vec u-[\xi]) e^{-2\sum_k \frac{u_k}{k} \xi^{-k}} 
\eeq

\item 
and
\bea\label{eq:equivH2}
& & \forall \ell\geq 1, \ \forall \alpha_1,\dots,\alpha_{\ell} \text{ such that } \sum_{i=1}^\ell\alpha_i=-1 \cr
 & 0 & =\Res_{\xi\to 0}\frac{d\xi}{\xi^2}  \hat \Tau(\vec t+[D]+[\xi]) \hat \Tau(\vec t-[D]-[\xi]) \prod_{j=1}^\ell (1-z_j/\xi)^{2\alpha_j} \cr
 & & \quad \text{in } \CC[[z_1,\dots,z_\ell]] 
 \text{ where } D = \sum_{j=1}^\ell \alpha_j.z_j \ .
\eea

\end{itemize}

In other words, Hirota equation is equivalent to \eqref{eq:equivH2}.

\et

\proof{

Start by assuming  \eqref{eq:equivH1}.
Let $D = \sum_{j=1}^\ell \alpha_j.z_j$ a divisor of degree $-1$.
We have
\beq
\hat \Tau(\vec t+[D]+[\xi]) \hat \Tau(\vec t-[D]-[\xi]) \prod_{j=1}^\ell (1-z_j/\xi)^{2\alpha_j} \in \CC[\xi,\xi^{-1}][[z_1,\dots,z_\ell]]
\eeq
and thus
\beq
\Res_{\xi\to 0}\frac{d\xi}{\xi^2}
\hat \Tau(\vec t+[D]+[\xi]) \hat \Tau(\vec t-[D]-[\xi]) \prod_{j=1}^\ell (1-z_j/\xi)^{2\alpha_j} \in \CC[[z_1,\dots,z_\ell]].
\eeq
In $\CC[[z_1,\dots,z_\ell]]$ we have $(1-z_j/\xi)^{2\alpha_j} = \exp(2\alpha_j \ln{(1-z_j/\xi)})$, and thus this can be rewritten
\beq
\Res_{\xi\to 0}\frac{d\xi}{\xi^2}
\hat \Tau(\vec t+[D]+[\xi]) \hat \Tau(\vec t-[D]-[\xi]) e^{-2\sum_{k\geq 1} \frac{\xi^{-k}}{k} \Tr D^k} \in \CC[[z_1,\dots,z_\ell]]
\eeq
and writting $u_k=\Tr D^k$, it can be rewritten
\beq
\Res_{\xi\to 0}\frac{d\xi}{\xi^2}
\hat \Tau(\vec t+\vec u+[\xi]) \hat \Tau(\vec t-\vec u-[\xi]) e^{-2\sum_k \frac{\xi^{-k}}{k} u_k} \in \CC[[u_1,u_2,\dots]].
\eeq
Thanks to \eqref{eq:equivH1}, it is vanishing, this implies \eqref{eq:equivH2}.

\smallskip

Now assume \eqref{eq:equivH2}.
Let $\mu=(\mu_1,\dots,\mu_\ell) $.
Our first step is to construct, for every $\ell$-upple $u_1,\dots,u_\ell$, a divisor $D$ of degree $-1$, such that for every $k\leq \ell$ we have $\Tr D^k=u_k$.

For that, let $N=2\ell+1$, let $\alpha_1=\dots=\alpha_\ell=1$ and $\alpha_{\ell+1}=\dots=\alpha_{2\ell+1}=-1$.
Let $z_{\ell+j} = 1$ for $j=1,\dots,\ell+1$.

Let
\bea
P(u_1,\dots,u_\ell;t) 
&=&  \sum_{m} \frac{(-1)^m}{m!} \sum_{k_1,\dots,k_m, \ k_i \geq 1, \ \sum_{i=1}^m k_i \leq \ell}  \prod_{i=1}^m \frac{(\ell+1) u_1^k +u_{k_i}}{k_i}   \ t^{\ell-\sum_{i=1}^m k_i} \cr
&=& \left[t^\ell \ e^{-\sum_{k=1}^\ell \frac{t^{-k}}{k} (u_k+(\ell+1)u_1^k)} \right]_+
\eea
where the second line means that we expand the exponential as a series of $1/t$, and keep only the non-negative powers of $t$.
$P(u_1,\dots,u_\ell;t)$ is thus a monic polynomial of $t$ of degree $\ell$.
It has $\ell$ zeros denoted $z_1,\dots,z_\ell$.
They are such that (this is the Newton's inversion formula)
\beq
\forall k=1,\dots,\ell
\qquad \sum_{i=1}^\ell z_i^k = (\ell+1)u_1^k + u_k.
\eeq
In other words, the divisor $D=\sum_{i=1}^\ell [z_i] - (\ell+1) [u_1]$ has degree $\deg D=-1$ and is such that 
\beq
\forall k=1,\dots,\ell
\qquad \Tr D^k = u_k.
\eeq
This defines a map:
\bea
\phi: & \CC^{\ell}  & \to \{\text{divisors of degree } -1\} \cr
& (u_1,\dots,u_\ell) & D = \sum_{i=1}^\ell [z_i] - (\ell+1) [u_1]
\eea
Moreover, $\phi$ respects the grading:
\beq
\phi(\lambda u_1 , \lambda^2 u_2,\dots,\lambda^{\ell} u_l) = \lambda \phi(u_1,u_2,\dots,u_\ell).
\eeq
By definition, we have (where $\phi(u_1,\dots,u_\ell)=D=\sum_{i=1}^\ell [z_i] - (\ell+1)[u_1]$):
\bea
&&\Res_{\xi\to 0}\frac{d\xi}{\xi^2} [\vec u^\mu] \hat \Tau(\vec t+\vec u+[\xi]) \hat \Tau(\vec t-\vec u-[\xi]) e^{-2\sum_k \frac{u_k}{k} \xi^{-k}} \cr
&&\Res_{\xi\to 0}\frac{d\xi}{\xi^2} [\vec u^\mu] \hat \Tau(\vec t+[\phi(\vec u)]+[\xi]) \hat \Tau(\vec t-[\phi(\vec u)]-[\xi]) \frac{\prod_{i=1}^{\ell} (1-z_j/\xi)^2}{(1-u_1/\xi)^{2(\ell+1)}}.   \cr
\eea
From \eqref{eq:equivH2} we know that
\beq
\Res_{\xi\to 0}\frac{d\xi}{\xi^2}  \hat \Tau(\vec t+[\phi(\vec u)]+[\xi]) \hat \Tau(\vec t-[\phi(\vec u)]-[\xi]) \frac{\prod_{i=1}^{\ell} (1-z_j/\xi)^2}{(1-u_1/\xi)^{2(\ell+1)}}   
= 0 \quad \text{in } \CC[[u_1,z_1,\dots,z_\ell]].
\eeq
This implies that it vanishes in $\CC[[u_1,u_2,\dots,u_\ell]]$,
this proves \eqref{eq:equivH1}.
}

With $\Tau$ this is reformulated as:
\bt[Hirota with divisors]
\label{th:HirotaDivisors}
$\Tau$ satisfies Hirota equation is equivalent to:

$\forall \ell\geq 1, \ \forall \alpha_1,\dots,\alpha_{\ell} \text{ such that } \sum_{i=1}^\ell\alpha_i=-1$
\beq\label{eq:equivH2Tau}
 0 =\Res_{\xi\to 0}   \Tau(\vec t+[D]+[\xi])  \Tau(\vec t-[D]-[\xi])   \quad \text{in } \CC[[z_1,\dots,z_\ell]] 
 \text{ where } D = \sum_{j=1}^\ell \alpha_j [z_j].
\eeq

\et

The following lemma to compute residues of formal series will be very useful
\bl[Residues of formal series]\label{lem:residuesformal}
If $f(z) = \sum_{j=0}^\infty \frac{f_j}{j!} z^j \in \CC[[z]] $ is a formal series, we have
\beq
\Res_0 \frac{d\xi}{\xi} f(\xi) \ (1-z/\xi)^{-1} = f(z) \quad \text{in } \CC[[z]] 
\eeq
and more generally
\beq
\Res_0 \frac{d\xi}{\xi^{n+1}} f(\xi) \ (1-z/\xi)^{-n-1} = \frac{1}{n!} f^{(n)}(z) \quad \text{in } \CC[[z]] 
\eeq
and this vanishes if $n<0$.

\el
In other words the formulas for residues of formal series are the same as if $f(z)$ were an analytic fuction in a neighborhood of  $z=0$.

\proof{
\bea
\Res_0 \frac{d\xi}{\xi^{n+1}} f(\xi) \ (1-z/\xi)^{-n-1} 
&=& \Res_0 \frac{d\xi}{\xi^{n+1}} \left( \sum_{j=0}^\infty \frac{f_j}{j!} \xi^j \right) \left( \sum_{k=0}^\infty \frac{(n+k)!}{k! n!} \frac{z^k}{\xi^k} \right)  \cr 
&=& \sum_{k=0}^\infty \frac{f_{n+k}}{(n+k)!} \frac{(n+k)!}{k! n!} z^k  \cr 
&=& \frac{1}{n!} \sum_{k=0}^\infty \frac{f_{n+k}}{k!}  z^k  \cr 
&=& \frac{1}{n!} f^{(n)}(z) \quad \text{in } \CC[[z]] .
\eea
}

\section{Fay Identities}
\label{sec:Fayid}
Let us apply the divisor formulation of Hirota in the case of a divisor with four points of the form
\beq
[D] = \frac12([z_1]-[z_2]-[\td z_1]-[\td z_2])
\eeq
(which is indeed of degree $\deg D=-1$).

\bt[Fay identities $n=2$]
\label{thm:Fay2}
Hirota equation implies:
\bea
&& \frac{(z_1-z_2)(\td z_1-\td z_2)}{(\td z_1-z_2)(\td z_2-z_2)(z_1-\td z_1)(z_1-\td z_2)} \ \frac{\hat\Tau(\vec{\td t} +[z_1]-[\td z_1]+[z_2]-[\td z_2])}{\hat\Tau(\vec{\td t})} \cr
&=&   \ \frac{\hat\Tau(\vec{\td t} +[z_1]-[\td z_2])}{(z_1-\td z_2)\hat\Tau(\vec{\td t})} \ 
\frac{\hat\Tau(\vec{\td t} +[z_2]-[\td z_1])}{(z_2-\td z_1)\hat\Tau(\vec{\td t})} 
 - \ \frac{\hat\Tau(\vec{\td t} +[z_1]-[\td z_1])}{(z_1-\td z_1)\hat\Tau(\vec{\td t})} \ \frac{\hat\Tau(\vec{\td t} +[z_2]-[\td z_2])}{(z_2-\td z_2)\hat\Tau(\vec{\td t})} \cr
\eea
or equivalently, with $\Tau$ instead of $\hat\Tau$:
\bea
&&  \frac{\Tau(\vec{\td t} +[z_1]-[\td z_1]+[z_2]-[\td z_2])}{\Tau(\vec{\td t})} \cr
&=&   \frac{\Tau(\vec{\td t} +[z_1]-[\td z_1])}{\Tau(\vec{\td t})} \ \frac{\Tau(\vec{\td t} +[z_2]-[\td z_2])}{\Tau(\vec{\td t})} \ - \frac{\Tau(\vec{\td t} +[z_1]-[\td z_2])}{\Tau(\vec{\td t})} \ 
\frac{\Tau(\vec{\td t} +[z_2]-[\td z_1])}{\Tau(\vec{\td t})} \  \cr
\eea
or equivalently, as a determinantal formula
\beq
 \frac{\Tau(\vec{\td t} +[z_1]-[\td z_1]+[z_2]-[\td z_2])}{\Tau(\vec{\td t})}
=   \det_{1\leq i,j\leq 2} \left(\frac{\Tau(\vec{\td t} +[z_i]-[\td z_j])}{\Tau(\vec{\td t})} \right) \  .
\eeq
All these equalities hold as formal series in $\CC[[z_1,z_2,\td z_1,\td z_2]]$.
\et

\proof{
Choose
\beq
[D] = \frac12([z_1]-[z_2]-[\td z_1]-[\td z_2]),
\eeq
and
\beq
\vec{\td t} = \vec t -[D]-[z_2].
\eeq
Hirota equation and \autoref{lem:residuesformal} imply
\bea
0&=& \Res_{} \frac{(z_1-\xi)d\xi}{(z_2-\xi)(\td z_1-\xi)(\td z_2-\xi)} \ \hat\Tau(\vec{\td t} +[z_1]-[\td z_1]-[\td z_2]+[\xi])
\hat\Tau(\vec{\td t} +[z_2]-[\xi]) \cr
&=& \frac{(z_1-z_2)}{(\td z_1-z_2)(\td z_2-z_2)} \ \hat\Tau(\vec{\td t} +[z_1]-[\td z_1]+[z_2]-[\td z_2]) \hat\Tau(\vec{\td t}) \cr
&& + \frac{(z_1-\td z_1)}{(z_2-\td z_1)(\td z_2-\td z_1)} \ \hat\Tau(\vec{\td t} +[z_1]-[\td z_2])
\hat\Tau(\vec{\td t} +[z_2]-[\td z_1]) \cr
&& + \frac{(z_1-\td z_2)}{(z_2-\td z_2)(\td z_1-\td z_2)} \ \hat\Tau(\vec{\td t} +[z_1]-[\td z_1])
\hat\Tau(\vec{\td t} +[z_2]-[\td z_2]) \cr
\eea
}

\bd[Kernel]
We define:
\beq
K(\xi,\xi') := \Tau(\vec t+[\xi']-[\xi])/{\Tau(\vec t)}
\eeq
It is a bi-spinor, it behaves near coinciding points as
\beq
K(\xi,\xi') = \frac{\sqrt{d\xi} \otimes \sqrt{d\xi'}}{\xi'-\xi}  \ (1+O(\xi'-\xi)).
\eeq
\ed

\bt[Fay identities for all $n\geq 1$]
\label{thm:Fay}

Let $D=\sum_{i=1}^n [z_i]-[\td z_i]$ a supersymmetric divisor (a unitary divisor of degree 0).
We have the determinantal formula:
\beq
 \frac{\Tau(\vec{\td t} +[D])}{\Tau(\vec{\td t})}
=   \det_{1\leq i,j\leq n} \left(\frac{\Tau(\vec{\td t} +[z_j]-[\td z_i])}{\Tau(\vec{\td t})} \right) \ = \det K(\td \xi_i,\xi_j)  
\eeq
as formal series in $\CC[[z_1,\dots,z_n,\td z_1,\dots,\td z_n ]]$.
\et
\proof{
We shall proceed by recursion on $n$. We know that this holds for $n=1$ and  $n=2$. Let $n\geq 3$.
Let us write:
\beq
D'_i=\sum_{j=2}^n [z_j]-\sum_{j\neq i} [\td z_j] = D-[z_1]+[\td z_i].
\eeq
Hirota equation gives:
\bea
0
&=& \Res_{0} \frac{d\xi}{z_1-\xi} \frac{\prod_{i=2}^n (\xi-z_i)}{\prod_{i=1}^n (\xi-\td z_i)} 
\ \hat \Tau(\vec t + [z_1]-[\xi]) \hat \Tau(\vec t - [\td z_1]+[\xi] + [D'_1]) \cr
&=& - \ \frac{\prod_{i=2}^n (z_1-z_i)}{\prod_{i=1}^n (z_1-\td z_i)}  
\ \hat \Tau(\vec t ) \hat \Tau(\vec t +[D]) \cr
&& + \sum_{i=1}^n  \frac{1}{z_1-\td z_i} \frac{\prod_{j=2}^n (\td z_i-z_j)}{\prod_{j\neq i}^n (\td z_i-\td z_j)}  \ 
\ \hat \Tau(\vec t +[z_1]-[\td z_i]) \hat \Tau(\vec t +[D'_i]) .
\eea
This gives
\bea
&&   \hat \Tau(\vec t ) \hat \Tau(\vec t +[D]) \cr
&=& \frac{\prod_{j=1}^n (z_1-\td z_j)}{\prod_{j=2}^n (z_1-z_j)}   \sum_{i=1}^n  \frac{1}{z_1-\td z_i} \frac{\prod_{j=2}^n (\td z_i-z_j)}{\prod_{j\neq i}^n (\td z_i-\td z_j)}  \ 
\ \hat \Tau(\vec t +[z_1]-[\td z_i]) \hat \Tau(\vec t +[D'_i]) .\cr
\eea
or equivalently in terms of $\Tau$:
\beq
  \frac{\Tau(\vec t +[D])}{\Tau(\vec t ) } 
= \sum_{i=1}^n (-1)^{i-1} \ \frac{\Tau(\vec t +[z_1]-[\td z_i])}{\Tau(\vec t ) } \ \frac{ \Tau(\vec t +[D'_i])}{\Tau(\vec t ) } .
\eeq

We compute the last factor with the recursion hypothesis:
\bea
  \frac{\Tau(\vec t +[D])}{\Tau(\vec t ) } 
&=& \sum_{i=1}^n  \frac{\Tau(\vec t +[z_1]-[\td z_i]) }{\Tau(\vec t ) }
\sum_{\sigma\in \mathfrak S_{n} \ , \ \sigma(1)=i} (-1)^\sigma \ \prod_{j=2}^n \frac{\Tau(\vec t +[z_j]-[\td z_{\sigma(j)}])}{\Tau(\vec t ) } \cr
&=& \sum_{\sigma\in \mathfrak S_{n} } (-1)^\sigma \ \prod_{j=1}^n \frac{\Tau(\vec t +[z_j]-[\td z_{\sigma(j)}])}{\Tau(\vec t ) } .
\eea
}

\bt[Fay $\implies$ Hirota]
\label{th:HirotaFay}
If $\Tau$ is a functional as in \autoref{def:Tauts}, that satisfies 
Fay determinantal formula for all $n$ and all divisors, then it satisfies Hirota.
\et

\proof{
(see \cite{GEKHTMAN200780})
Assume that Fay identities hold for all $n\geq 2$, i.e. all unitary divisor of degree 0. 

For $n=2$, choosing 
\beq
[D] = \frac12([z_1]-[z_2]-[\td z_1]-[\td z_2])
\eeq
and
\beq
\vec{\td t} = \vec t -[D]-[z_2].
\eeq
We find
\bea
&& \Res_{0} \frac{(z_1-\xi)d\xi}{(z_2-\xi)(\td z_1-\xi)(\td z_2-\xi)} \ \hat\Tau(\vec{\td t} +[z_1]-[\td z_1]-[\td z_2]+[\xi])
\hat\Tau(\vec{\td t} +[z_2]-[\xi]) \cr
&=& \frac{(z_1-z_2)}{(\td z_1-z_2)(\td z_2-z_2)} \ \hat\Tau(\vec{\td t} +[z_1]-[\td z_1]+[z_2]-[\td z_2]) \hat\Tau(\vec{\td t}) \cr
&& + \frac{(z_1-\td z_1)}{(z_2-\td z_1)(\td z_2-\td z_1)} \ \hat\Tau(\vec{\td t} +[z_1]-[\td z_2])
\hat\Tau(\vec{\td t} +[z_2]-[\td z_1]) \cr
&& + \frac{(z_1-\td z_2)}{(z_2-\td z_2)(\td z_1-\td z_2)} \ \hat\Tau(\vec{\td t} +[z_1]-[\td z_1])
\hat\Tau(\vec{\td t} +[z_2]-[\td z_2]) \cr
&=& 0.
\eea
Then, by a recursion on $n$ (the same as in the proof of \autoref{thm:Fay} above), we find
\bea
&& \Res_{0} \frac{d\xi}{z_1-\xi} \frac{\prod_{i=2}^n (\xi-z_i)}{\prod_{i=1}^n (\xi-\td z_i)} 
\ \frac{\hat \Tau(\vec t + [z_1]-[\xi])}{\Tau(\vec t)} \frac{\hat \Tau(\vec t - [\td z_1]+[\xi] + [D'_1])}{\Tau(\vec t)} \cr
&\propto & 
\Big(
\frac{\Tau(\vec t +[D])}{\Tau(\vec t)}  
 - \sum_\sigma (-1)^\sigma \prod_{i=1}^n K(\td z_i, z_{\sigma(i)})\Big) \cr
 &=& 0.
\eea
Since this holds for all $n$ and all $D$, this proves Hirota.
}

\bt[Reproducing kernel]
$K(\xi,\xi')$ is a self--reproducing kernel for the insertion operator:
\beq
\Delta_\xi \left(\frac{ \Tau(\vec t+[\xi']-[\xi''])}{\Tau(\vec t)}\right) =  - \ \frac{ \Tau(\vec t+[\xi']-[\xi])}{\Tau(\vec t)} \ \frac{ \Tau(\vec t+[\xi]-[\xi''])}{\Tau(\vec t)}.
\eeq
or equivalently
\beq
\Delta_\xi K(\xi',\xi'') = - K(\xi',\xi)K(\xi,\xi'').
\eeq
\et

\proof{

Take $D=\frac12([\xi_1]-[\xi_2])-[\xi_0]$ in \eqref{eq:equivH2} and $\vec{\td t}=\vec t + \frac12([\xi_1]-[\xi_2]) $
\bea
0 
& = & \Res_0 \frac{(\xi-\xi_1)d\xi}{(\xi-\xi_2)(\xi-\xi_0)^2}
\hat\Tau(\vec {\td t}+[\xi]+\frac12([\xi_1]-[\xi_2])-[\xi_0])\hat\Tau(\vec {\td t}-[\xi]-\frac12([\xi_1]-[\xi_2])+[\xi_0]) \cr
& = & \Res_0 \frac{(\xi-\xi_1)d\xi}{(\xi-\xi_2)(\xi-\xi_0)^2}
\hat\Tau(\vec t+[\xi]+[\xi_1]-[\xi_2]-[\xi_0])\hat\Tau(\vec t-[\xi]+[\xi_0]) \cr
& = & \frac{(\xi_2-\xi_1)}{(\xi_2-\xi_0)^2}
\hat\Tau(\vec t+[\xi_1]-[\xi_0])\hat\Tau(\vec t-[\xi_2]+[\xi_0]) \cr
&  & + \frac{d}{d\xi} \left( \frac{(\xi-\xi_1)}{(\xi-\xi_2)}
\hat\Tau(\vec t+[\xi]+[\xi_1]-[\xi_2]-[\xi_0])\hat\Tau(\vec t-[\xi]+[\xi_0]) \right)_{\xi=\xi_0} \cr
& = & \frac{(\xi_2-\xi_1)}{(\xi_2-\xi_0)^2}
\hat\Tau(\vec t+[\xi_1]-[\xi_0])\hat\Tau(\vec t-[\xi_2]+[\xi_0]) \cr
&  & + \frac{1}{(\xi_0-\xi_2)}
\hat\Tau(\vec t+[\xi_1]-[\xi_2])\hat\Tau(\vec t) 
 - \frac{(\xi_0-\xi_1)}{(\xi_0-\xi_2)^2}
\hat\Tau(\vec t+[\xi_1]-[\xi_2])\hat\Tau(\vec t) \cr
&  & + \frac{(\xi_0-\xi_1)}{(\xi_0-\xi_2)} \frac{d}{d\xi} \left( 
\hat\Tau(\vec t+[\xi]+[\xi_1]-[\xi_2]-[\xi_0])\right)_{\xi=\xi_0}\hat\Tau(\vec t-[\xi]+[\xi_0])  \cr
&  & + \frac{(\xi_0-\xi_1)}{(\xi_0-\xi_2)} 
\hat\Tau(\vec t+[\xi]+[\xi_1]-[\xi_2]-[\xi_0])\frac{d}{d\xi} \left( \hat\Tau(\vec t-[\xi]+[\xi_0]) \right)_{\xi=\xi_0} \cr
& = & \frac{(\xi_2-\xi_1)}{(\xi_2-\xi_0)^2}
\hat\Tau(\vec t+[\xi_1]-[\xi_0])\hat\Tau(\vec t-[\xi_2]+[\xi_0]) \cr
&  & + \frac{\xi_1-\xi_2}{(\xi_0-\xi_2)^2}
\hat\Tau(\vec t+[\xi_1]-[\xi_2])\hat\Tau(\vec t)  \cr
&  & + \frac{(\xi_0-\xi_1)}{(\xi_0-\xi_2)} \Delta_{\xi_0}
\hat\Tau(\vec t+[\xi_1]-[\xi_2]) \ \hat\Tau(\vec t)  \cr
&  & - \frac{(\xi_0-\xi_1)}{(\xi_0-\xi_2)} 
\hat\Tau(\vec t+[\xi_1]-[\xi_2])\ \Delta_{\xi_0} \hat\Tau(\vec t).  \cr
\eea
This implies
\bea
\frac{(\xi_2-\xi_1)}{(\xi_0-\xi_1)(\xi_2-\xi_0)}\frac{\hat\Tau(\vec t+[\xi_1]-[\xi_0])}{\hat\Tau(\vec t)} \ \frac{\hat\Tau(\vec t-[\xi_2]+[\xi_0])}{\hat\Tau(\vec t)} 
& = &  \frac{\xi_1-\xi_2}{(\xi_0-\xi_2)(\xi_0-\xi_1)}
\frac{\hat\Tau(\vec t+[\xi_1]-[\xi_2])}{\hat\Tau(\vec t)}   \cr
&  & + \Delta_{\xi_0} \left( \frac{\hat\Tau(\vec t+[\xi_1]-[\xi_2])}{\hat\Tau(\vec t)}\right),  \cr
\eea
i.e.
\bea
\frac{\Tau(\vec t+[\xi_1]-[\xi_0])}{\Tau(\vec t)} \ \frac{\Tau(\vec t-[\xi_2]+[\xi_0])}{\Tau(\vec t)} 
& =  & - \Delta_{\xi_0} \left( \frac{\Tau(\vec t+[\xi_1]-[\xi_2])}{\Tau(\vec t)}\right),  \cr
\eea
i.e.
\beq
\Delta_{\xi_0} K(\xi_2,\xi_1) = - K(\xi_2,\xi_0) K(\xi_0,\xi_1).
\eeq
}

\bt[determinantal formulas for $W_n$]
\beq
W_1(\xi) = \lim_{\xi'\to \xi} K(\xi,\xi') - \frac{1}{E(\xi',\xi)}.
\eeq
For $n\geq 2$
\beq
W_n(\xi_1,\dots,\xi_n) 
= \sum_{\sigma \in \mathfrak S_n^{1-\text{cycle}}} (-1)^\sigma
\prod_{i=1}^N \frac{\Tau(\vec t + [\xi_i]-[\xi_{\sigma(i)}])}{\Tau(\vec t)}
=  \sum_{\sigma \in \mathfrak S_n^{1-\text{cycle}}} (-1)^\sigma
\prod_{i=1}^N K(\xi_{\sigma(i)},\xi_i) .
\eeq

\et

\proof{

For $n=1$, we have from \eqref{eq:shiftexpDelta} of \autoref{lemma:derivTauWn}:
\bea
\lim_{\xi'\to \xi} K(\xi,\xi') - \frac{1}{E(\xi',\xi)}
&=& \lim_{\xi'\to \xi} \frac{\hat\Tau(\vec t+[\xi']-[\xi])}{E(\xi',\xi)\hat\Tau(\vec t)} - \frac{1}{E(\xi',\xi)}  \cr
&=& \lim_{\xi'\to \xi} \frac{\hat\Tau(\vec t+[\xi']-[\xi])-\hat\Tau(\vec t)}{E(\xi',\xi)\hat\Tau(\vec t)}  \cr
&=& \frac{d\xi}{\hat\Tau(\vec t)} \frac{d}{d\xi'}\left( \hat\Tau(\vec t+[\xi']-[\xi])\right)_{\xi'=\xi} \cr
&=& \frac{1}{\hat\Tau(\vec t)} \Delta_\xi \hat\Tau(\vec t) \cr
&=&  \Delta_\xi \ln\hat\Tau(\vec t) \cr
&=& W_1(\xi).
\eea

Then act with $\Delta_{\xi_2}$:
\bea
W_2(\xi_1,\xi_2) &=&
\Delta_{\xi_2} W_1(\xi_1) \cr
&=& \Delta_{\xi_2} \left( \lim_{\xi'\to \xi_1} K(\xi_1,\xi') - \frac{1}{E(\xi',\xi_1)}\right) \cr
&=&  \lim_{\xi'\to \xi_1} -K(\xi_1,\xi_2)K(\xi_2,\xi') \cr
&=&  -K(\xi_1,\xi_2)K(\xi_2,\xi_1) .
\eea
Then act recursively with $\Delta_{\xi_n}$:
for $n+1\geq 3$, assume it is proved for $n$:
\bea
W_{n+1}(\xi_1,\dots,\xi_n,\xi_{n+1})
&=& \Delta_{\xi_{n+1}}  W_{n}(\xi_1,\dots,\xi_n) \cr
&=& \Delta_{\xi_{n+1}} \sum_{\sigma\in\mathfrak S_n} (-1)^{n-1} \prod_{i=1}^n K(x_i,x_{\sigma(i)}) \cr
&=&  \sum_{\sigma\in\mathfrak S_n^{1-\text{cycle}}} (-1)^{n-1} \sum_{j=1}^n \Delta_{\xi_{n+1}}K(x_j,x_{\sigma(j)})  \prod_{i=1, \ i\neq j}^n K(x_i,x_{\sigma(i)}) \cr
&=&  \sum_{\sigma\in\mathfrak S_n^{1-\text{cycle}}} (-1)^{n} \sum_{j=1}^n K(x_j,x_{n+1})K(x_{n+1},x_{\sigma(j)})  \prod_{i=1, \ i\neq j}^n K(x_i,x_{\sigma(i)}) \cr
&=& \sum_{j=1}^n \sum_{\sigma\in\mathfrak S_{n+1}^{1-\text{cycle}}, \ \sigma(j)=n+1} (-1)^{n}  \prod_{i=1}^{n+1} K(x_i,x_{\sigma(i)}) \cr
&=& \sum_{\sigma\in\mathfrak S_{n+1}^{1-\text{cycle}}} (-1)^{n}  \prod_{i=1}^{n+1} K(x_i,x_{\sigma(i)}) \cr
\eea
}

\subsection{Miwa Jimbo}

\bt[Miwa Jimbo]
\beq
\frac{\partial}{\partial t_k}
\ln\Tau(\vec t) = \Res_{\xi\to 0} \frac{1}{k} \xi^{-k} W_1(\xi).
\eeq
\beq
\frac{\partial^n}{\partial t_{k_1}\dots \partial t_{k_n}}
\ln\Tau(\vec t) = \Res_{\xi_1\to 0}\dots\Res_{\xi_n\to 0} \frac{\xi_1^{-k_1}\dots \xi_n^{-k_n}}{k_1\dots k_n}  W_n(\xi_1,\dots,\xi_n).
\eeq
\et

\proof{Use eq~\eqref{eqdef:InsertionOp} and eq~\eqref{eq:Deltashiftcorr}.}

\subsection{Conclusion on formal $\Tau$-functions}

Here we recalled how Hirota equations are equivalent to Fay identities in the case of formal series times exponentials.

Let us summarize the main points:

\begin{itemize}

\item $\Tau$ function here was a formal series of times, multiplied by an exponential factor, that is a single trans-series, or also called a trans-monomial. 

\item Hirota equations are an infinite set of compatible partial differential equations. We insist that this requires infinitely many variable times: the Tau function has to be defined on an Abelian group $\Gamma$ (which is a group acting on a Grassmannian in the Sato Segal-Wilson formalism), or more generally on an infinite dimensional space for which times are local coordinates.

\item Fay identities, are expressed as relations between function of $2n$ complex variables $(\xi_i, \td \xi_i)_{i=1,\dots n} $ rather than times. Somehow we have traded an infinite-time-dependence to functions of a finite number of complex variables.

The points $\xi_1,\dots,\xi_n,\td\xi_1,\dots,\td\xi_n $ are  complex variables in which we take residues. Residue is a notion attached to complex variables, i.e. living on a Riemann surface.

\item This justifies that we are going to extend the present formalism to the case where $\xi_i, \td\xi_i$s belong to a Riemann surface. And Fay identities are then functional relations among functions on a Riemann surface at fixed times. This doesn't need to have a ``space" of times, and this doesn't need series expansions.

\item We are thus going to enlarge the context of Fay identities to genuine functions on a Riemann surface, rather than series, trans-monomials or transseries (trans-series are linear combinations of trans-monomials).

\item In some sense it is wrong to say that Hirota $\Leftrightarrow$ Fay, because Fay is more general, it makes sense even when we don't have series times exponentials. Hirota is somehow only the Taylor expansion, i.e. the germ of Fay.

\end{itemize}



\section{Tau functions and Riemann Surfaces}
\label{sec:TauRS}
\subsection{Introduction and motivation}

The goal is to exhibit some solutions of Fay identities.
This means exhibiting a space $\Gamma$ with some coordinates $\vec t$, and some function $\Tau$ on $\Gamma$ satisfying Hirota equations, or equivalently Fay identities.

\smallskip

Fay identities are relations among $\Tau$ evaluated at times Sato--shifted by some $\xi_i$s.
Each $\xi_i$ is a complex variable, and can be viewed as a local coordinate in a chart  of a Riemann surface, so that $\Tau(\vec t+ [\xi])$ is an analytic function on the Riemann surface.

Punctures are defined as the places where $\Tau(\vec t+[\xi])$ is not analytic, this is related to the exponentials of negative powers.
This means by definition, that $\Tau(\vec t+[\xi])$ should be analytic elsewhere, on the whole pointed surface without the punctures. So it is natural to associate a Tau functions to Riemann surfaces. There are several classical ways of associating a space $\Gamma$ to Riemann surfaces, and finding a Tau function on it.

\medskip

\textbf{Solutions presented here:}

In each case, the ``times" are local coordinates in a space $\Gamma$.

\begin{enumerate}

\item \textbf{Isospectral Tau functions (\autoref{sec:Tau1} below):}  here we shall consider $\Gamma$ to be the space of meromorphic 1-forms on a fixed Riemann surface $\curve$. 
The times are ``moments": if $\Omega\in\Gamma$ is a meromorphic 1-form times are $t_{k}(\Omega) = \underset{0}{\Res} \xi^k \Omega$. 
The Sato vector $[\xi]$ corresponds to a meromorphic 1-form $\Omega_\xi$ having a simple pole at position $\xi$, with residue 1. For a divisor $D$, the Sato vector corresponds to a meromorphic 1-form $\Omega_D$ with simple poles at $\operatorname{supp} D$ with residues = weights. Fay identities are then relations between analytic functions on the Riemann surface.
The solutions of Fay identities in this case are known explicitely, they are expressed with Theta-functions.
This is also called the ``reconstruction formula'' or ``finite-gap solution", explicited in \autoref{sec:Tau1} below.

\item \textbf{Isomonodromic Tau functions (\autoref{sec:Tau2} below):} Here we shall consider $\Gamma$ to be a moduli space of ``spectral curves". A spectral curve is a Riemann surface $\curve$ equipped with an immersion into a cotangent space $\curve\hookrightarrow T^*\curverond$, $z\mapsto (x,y)$. 
Times are moments $t_{k} = \underset{0}{\Res} \xi^k y dx $. 
A Sato-shifted spectral curve by a Sato vector $[\xi]$ corresponds to a  spectral curve whose immersion $y$ has a simple pole at position $\xi$ with residue 1, and more generally for a divisor $D$ the Sato-shifted spectral curve by the Sato vector $[D]$ corresponds to a  spectral curve whose immersion $y$ has poles at $\operatorname{supp} D$ with residues = weights. 

The solutions of Fay identities in this case are called ``isomonodromic Tau functions".

Small deformations of spectral curves are meromorphic 1-forms $\Omega= \delta(ydx)$, i.e. the tangent space of the moduli space of spectral curve is the space of meromorphic 1-forms.
This means  that isospectral Tau functions can be recovered as an ``infinitesimal" limit  of isomonodromic Tau functions.

This Tau function of spectral curves is developed in \autoref{sec:Tau2} below.

\item Many other constructions exist, and we do not claim to be exhaustive here. An example is with $\Gamma$  the deRham space of connections of principal bundles of reductive Lie group $G$ over a Riemann surface $\curverond$. This is related to Hitchin systems.
These constructions are not presented here.

In \autoref{sec:matrices} we present a matrix integral solution of Fay identities.

\end{enumerate}

\subsection{Riemann surfaces}

Let us first introduce some geometry of Riemann surfaces.

Let $\curverond=\CC P^1$, we call it the ``base" curve. The base curve may be any Riemann surface but, in this section, for simplicity we restrict to  $\curverond = $ Riemann sphere  = $\CC P^1$ = $\CC \cup\{\infty\}$.
\bd[Ramified cover]
Let $\curve$ a compact Riemann surface of some genus $\genus$, equipped with a holomorphic ramified covering map
\beq
\x: \curve \to \curverond.
\eeq
\ed

\bd[Canonical local coordinates]
At each point $p$ of $\curve$, we define $x_p=\x(p)$ if $\x(p)\neq \infty$ and $x_p=0$ if $\x(p)=\infty$. 
Let
\beq
a_p := \idx_p (\x-x_p).
\eeq

If $\x(p)\neq \infty$, we have $a_p>0$, and the canonical local coordinate is $\xi_p=(\x-x_p)^{\frac{1}{a_p}}$.

If $\x(p)=\infty$, we have $a_p<0$, and the canonical local coordinate is $\xi_p=\x^{\frac{1}{a_p}} = \x^{-\frac{1}{|a_p|}}$. 

In both cases
\beq
\x = x_p+\xi_p^{a_p}.
\eeq

Remark that if $p$ is a ramification point, we have $|a_p|\geq 2$, and thus $\xi_p = (\x-x_p)^{1/a_p}$ is ambiguous.
It is defined up to a root of unity, or alternatively, there are $|a_p|$ possible different choices for $\xi_p$:
\beq
\xi_{p,(k)} = \rho_{a_p}^k \xi_p
\ , \ k=1,\dots,|a_p| \ , 
\qquad \text{where} \ \rho_n = e^{\frac{1}{n}2\pi i}.
\eeq
\ed

\bd
Let
\beq
\mathfrak M^1(\curve)
\eeq
the complex vector space of meromorphic 1-forms on $\curve$.
It is an infinite dimensional complex vector space.

\ed

We associate to a meromorphic 1-form $\Omega$ on $\curve$ a family of times.
\bd[Times of a meromorphic 1-form]
Let $\Omega\in \mathfrak M^1(\curve) $ a meromorphic 1-form on $\curve$.
Since $\curve$ is compact, $\Omega$ has a finite number of poles.

For a pole $p$ of $\Omega$, we can write its polar part as
\bea
\Omega 
&=& \sum_{k=0}^{-1+\deg_p\Omega} t_{p,k}(\Omega) \xi_p^{-k-1}d\xi_p + \text{holomorphic at } p \cr
&=& \frac{1}{a_p}\sum_{k=0}^{-1+\deg_p\Omega} t_{p,k}(\Omega) (\x-x_p)^{-\frac{k}{a_p}-1}d\x + \text{holomorphic at } p.
\eea
The coefficients $t_{p,k}(\Omega)$ of the polar part are worth
\beq
t_{p,k}(\Omega) = \Res_p \xi_p^k \ \Omega .
\eeq
We define the infinite vectors
\beq
\vec t_{p}(\Omega) := (t_{p,1}(\Omega),t_{p,2}(\Omega),\dots,t_{p,-1+\deg_p\Omega}(\Omega),0,0,\dots).
\eeq
Since $\Omega$ is meromorphic, there can be only a finite number of non-vanishing times.

We write collectively:
\beq
\mathbf t(\Omega) =  \{\vec t_p(\Omega)\}_{p=\text{poles of }\Omega}.
\eeq

Remark that if $p$ is at the same time a pole and a ramification point, we have several families of times, related by a root of unity:
\beq
\vec t_{p,(k)}(\Omega) = (\rho_{a_p}^k t_{p,1}(\Omega), \rho_{a_p}^{2k} t_{p,2}(\Omega), \dots, \rho_{a_p}^{jk} t_{p,j}(\Omega),   \dots) := \rho_{a_p}^k(\vec t_p(\Omega)).
\eeq

Remark that if $p$ is not a pole of $\Omega$ we have $\vec t_p(\Omega)=0$.

\ed

\subsubsection{Monodromies and periods}

The curve $\curve$ has some genus $\genus$.
It means that there are non-contractible cycles, the fundamental group $\pi_1(\curve)$ has rank $2\genus$.

\bd[Marking]
It is possible (not unique) to find a basis of $2\genus$ Jordan loops, $(\acycle_j,\bcycle_j)_{j=1,\dots,\genus}$ that generate $\pi_1(\curve)$, and that have transverse symplectic intersections:
\beq
\acycle_i \cap \bcycle_j = \delta_{i,j}
\quad , \quad
\acycle_i \cap \acycle_j = 0
\quad , \quad
\bcycle_i \cap \bcycle_j = 0,
\eeq
for all $i,j$ in $[1,\dots,\genus]$.
\ed
\br
 We choose Jordan loops rather than cycles (cycles are homotopy classes of Jordan loops and their linear combinations), because the exact support will matter, since we will integrate meromorphic forms that can have poles.
\er

\bd[Periods of a meromorphic 1-form]
Let $\Omega\in \mathfrak M^1(\curve) $ a meromorphic 1-form.
Let us assume that the poles of $\Omega$ are not on the support of our marking symplectic Jordan loops.
We define the ``periods" of $\Omega$ as follows:
for all $j\in [1,\dots,\genus]$
\beq
\epsilon_j(\Omega) := \frac{1}{2\pi i} \oint_{\acycle_j} \Omega.
\eeq
Periods are also called 1st kind times.
\ed

\br
If $\Omega$ would have poles on the support of the Jordan loop, one should replace the integral $\oint_{\acycle_j}\Omega$ by the regularized integral defined in \cite{EynCycles}, so that all what follows below remains valid for all $\Omega$.
But here, to simplify, let us focus on the case where $\Omega$ has no pole on the marked Jordan loops, and regularized integrals are just usual integrals.
\er

\bd[3rd kind times]
For each pole $p$ of $\Omega$, let
\beq
t_{p,0}(\Omega) = \Res_p \Omega.
\eeq

\ed

\subsubsection{Canonical decomposition of meromorphic forms}

\bl[1st kind differentials]
\label{def:hol1forms}

For each $i=1,\dots,\genus$, there exists a unique holomorphic 1-form on $\curve$, denoted $\omega'_{i}$, with no pole,
and such that
\beq
\forall \ j=1,\dots,\genus \ , \qquad
\oint_{\acycle_j} \omega'_{i}=\delta_{i,j}.
\eeq
$\omega'_1,\dots,\omega'_\genus$ form a basis of $\Omega^1(\curve)$, the space of holomorphic 1-forms on $\curve$, it is called the basis dual to the marking symplectic Jordan loops.
\el
\proof{This is a classical theorem of Riemann surfaces, see \cite{Farkas}}.

\bl[3rd kind differentials]
\label{lemdef:3rdkinddif}

Let $p,q$ two points on $\curve$ (we assume that they are not on the support of the marking Jordan loops).

There exists a unique meromorphic 1-form, denoted $\omega'''_{p,q}$, that has a simple pole at $p$ of residue $+1$ and a simple pole at $q$ of residue $-1$ and no other pole.
\beq
\Res_p \omega'''_{p,q}= 1 = - \Res_q \omega'''_{p,q}
\eeq
and normalized on $\acycle$-cycles
\beq
\forall \ i=1,\dots,\genus \ , \qquad
\oint_{\acycle_i} \omega'''_{p,q}=0.
\eeq
If $D=\sum_{i=1}^\ell \alpha_i.[z_i] $  is a divisor of degree $\deg D=0$, we define
\beq
\omega'''_D := \sum_{i=1}^\ell \alpha_i \omega'''_{z_i,o}
\eeq
where $o$ is a generic point of $\curve$, and in fact $\omega'''_D$ is independent of it.
Remark that $\omega'''_{p,q} = \omega'''_{[p]-[q]}$.

$\omega'''_D$ is the unique meromorphic 1-form that  has simple poles at the support of $D$ with residues = weights and no other poles, and vanishing $\acycle$-cycle integrals.

\el
\proof{This is a classical theorem, see \cite{Farkas}.}

\bl[2nd kind differentials]
Let $p$ a point on $\curve$, and $k\geq 1$ (we assume that $p$ is not on the support of the marking Jordan loops).

There exists a unique meromorphic 1-form, denoted $\omega''_{p,k}$, that has a pole at $p$  of degree $k+1$ and no other pole, and behaving like
\beq
\omega''_{p,k} = \xi_p^{-k-1}d\xi_{p} + \text{holomorphic at } p,
\eeq
and normalized on $\acycle$-cycles
\beq
\forall \ i=1,\dots,\genus \ , \qquad
\oint_{\acycle_i} \omega''_{p,k}=0.
\eeq

\el
\proof{This is a classical theorem, see \cite{Farkas}}

\bl[Decomposition]
Every meromorphic 1-form $\Omega$ can be uniquely written
\bea
\Omega 
&=& \sum_{p=\text{poles of }\Omega} \sum_{k=1}^{-1+\deg_p\Omega} t_{p,k}(\Omega) \ \omega''_{p,k} \cr
&& + \sum_{p=\text{poles of }\Omega}  t_{p,0}(\Omega) \ \omega'''_{p,o} \cr
&& + 2\pi\ii\sum_{i=1}^{\genus} \epsilon_i(\Omega) \ \omega'_i .
\eea
(here $o$ is an arbitrary generic point of $\curve$, and the decomposition is actually independent of it).

In other words the times $\mathbf t(\Omega)$ are the coordinates of $\Omega$ in the cannonical basis of forms.

\el

\proof{This is a classical theorem, see \cite{Farkas}.
The main idea is that the difference between $\Omega$ and the right hand side is a 1-form without poles, and with vanishing $\acycle_i$ integrals, so it must vanish.
}

\subsection{More geometry of Riemann surfaces}

Before showing explicit solutions of the Fay Hirota equations we first need to introduce more geometry, in particular, Theta functions and Fay's Prime form \cite{Farkas,Fay,TataLectures}.

\subsubsection{Holomorphic forms, Riemann periods, Theta}

\bp[Riemann]
Define the Riemann matrix of periods using the normalized holomorphic 1-forms of \autoref{def:hol1forms}
\beq
\tau_{i,j} = \oint_{\bcycle_i} \omega'_j.
\eeq
$\tau$ is a $\genus\times\genus$ Siegel matrix, which means it is symmetric and its imaginary part is positive definite:
\beq
\tau=\tau^t 
\quad , \quad
\Im\tau>0.
\eeq

\ep
\proof{This is a very classical theorem going back to Riemann \cite{Farkas}.}

\bd[Abel map]
Let $o$ a generic point of $\curve$ fixed once for all, called the origin.
We define the Abel map:
\bea
\curve^{\text{universal cover}} & \to & \CC^\genus \cr
z & \mapsto & \mathfrak a(z) = (\mathfrak a_1(z),\dots,\mathfrak a_\genus(z)) \qquad \text{where } \ \mathfrak a_i(z)= \int_o^z \omega'_i.
\eea
They have monodromies
\beq
\mathfrak a_i(z+\acycle_j) = \mathfrak a_i(z) + \delta_{i,j},
\eeq
\beq
\mathfrak a_i(z+\bcycle_j) = \mathfrak a_i(z) + \tau_{i,j}
\eeq
and therefore,
\bea
\curve & \to & \mathbb J = \CC^\genus/(\ZZ^\genus+\tau\ZZ^\genus) \cr
z & \mapsto & \mathfrak a(z) \ \mod \ZZ^\genus+\tau\ZZ^\genus
\eea
is well defined.
$\mathbb J = \CC^\genus/(\ZZ^\genus+\tau\ZZ^\genus)$ is called the Jacobian of $\curve$.

If $D=\sum_{i=1}^\ell \alpha_i.z_i$ is a divisor of points of $\curve$, we define by linearity
\beq
\mathfrak a(D) = \sum_{i=1}^\ell \alpha_i \mathfrak a(z_i).
\eeq
If $\deg D=0$, $\mathfrak a(D)$ is independent of the origin $o$.

\ed

\bd[Riemann Theta function]
Let $\tau$ a Siegel matrix.
We define the function
\bea
\CC^\genus & \to & \CC \cr
u=(u_1,\dots,u_\genus) & \mapsto & \Theta(u) = \Theta(u;\tau) =: \sum_{\mathbf n\in \ZZ^\genus}
e^{2\pi \ii (u,\mathbf n)} \ e^{\pi \ii (\mathbf n,\tau \mathbf n)}
\eea
(very often we shall consider $\tau$ fixed and view $\Theta$ as only a function of $u\in \CC^\genus$).
Thanks to $\Im\tau>0$, the series is absolutely convergent for all $u\in \CC^\genus$ and is an anlytic entire function on $\CC^\genus$.

\ed

\bp[Riemann]
$\Theta$ obeys the following properties:
\bea
\Theta(-u) & =& \Theta(u) \cr
\Theta(u+\mathbf n) &=& \Theta(u) \cr
\Theta(u+\tau\mathbf n) &=& \Theta(u) \ e^{-2\pi\ii (u,\mathbf n)} e^{-\pi\ii (\mathbf n,\tau\mathbf n)} \cr
\Theta(u+\frac12\mathbf n+\frac12 \tau \mathbf m) &=& 0 \qquad \text{iff} \ (\mathbf n,\mathbf m)\in 2\ZZ+1
\eea

\ep

\bd[Half-integer characteristic]
Let $\mathbf n,\mathbf m\in \ZZ^\genus$.
\beq
\chi := \frac12\mathbf n+\frac12 \tau \mathbf m
\eeq
is called a half-integer characteristics.

It is called odd iff $(\mathbf n,\mathbf m)\in 2\ZZ+1$.

It is called non-singular iff the vector $\Theta'(\chi) = (\Theta'_1(\chi),\dots,\Theta'_\genus(\chi))\neq 0$.

It is known that there always exists some non-singular odd half-integer characteristic.

From now on, let us choose one of them, fixed once for all.
Let us define the following holomorphic 1-form
\beq
h_\chi := \sum_{i=1}^\genus \Theta'_i(\chi) \omega_i.
\eeq
It is well known that $h_\chi$ is a holomorphic 1-form, that has $\genus-1$ double zeros on $\curve$.

\ed

\subsubsection{Fay's prime forms}

\bd[Fay's prime form]
Let
\beq
E(z_1,z_2) := \frac{\Theta(\mathfrak a(z_1)-\mathfrak a(z_2)+
\chi)}{\sqrt{h_\chi(z_1) \ h_\chi(z_2)}}.
\eeq
If $D =\sum_{i=1}^\ell \alpha_i.z_i $ is a divisor, we define
\beq
E(D) := \prod_{i<j} E(z_i,z_j)^{-\alpha_i\alpha_j}.
\eeq
Remark:
\beq
E(z_1,z_2) = E([z_1]-[z_2]).
\eeq

\ed

\bp[Fay]
$E(z_1,z_2)$ is locally analytic of $z_1$ (resp. $z_2$), or more precisely is analytic on the universal cover. In particular it has no poles (the $\genus-1$ doubles zeros of $h_\chi$ happen to be exactly $\genus-1$ zeros of $\Theta$ in the numerator).
It vanishes only at $z_1=z_2$ and vanishes linearly.
In any local coordinate $\xi_i=\xi(z_i)$ it behaves as
\beq
E(z_1,z_2) \mathop{\sim}_{z_1\to z_2} \frac{\xi_1-\xi_2}{\sqrt{d\xi_1 d\xi_2}} \ (1+O(\xi_1-\xi_2)).
\eeq

\ep

\bd[Szeg\"o kernel]
Let $\Omega$ a meromorphic 1-form on $\curve$.
Let 
\beq
\zeta(\Omega)=(\zeta_1(\Omega),\dots,\zeta_\genus(\Omega)) := \frac{1}{2\pi\ii}\oint_{\bcycle-\tau\acycle} \Omega,
\eeq
\beq
\text{i.e.} \qquad \zeta_{i}(\Omega) = \frac{1}{2\pi\ii} \left( \oint_{\bcycle_i}\Omega - \sum_{j=1}^\genus \tau_{i,j} \oint_{\acycle_j} \Omega \right).
\eeq
Let the Szeg\"o kernel
\beq
\psi([z_1]-[z_2];\Omega) := \frac{\Theta(\zeta(\Omega)+\mathfrak a(z_1)-\mathfrak a(z_2)+\chi)}{E(z_1,z_2)\Theta(\zeta(\Omega)+\chi) } \ e^{\int_{z_2}^{z_1} \Omega}\ e^{-2\pi\ii (\epsilon(\Omega),\mathfrak a(z_1)-\mathfrak a(z_2))}.
\eeq
More generally, for a divisor $D$ of degree 0:
\beq
\psi(D;\Omega) =
\frac{\Theta(\zeta(\Omega)+\mathfrak a(D)+\chi)}{E(D) \ \Theta(\zeta(\Omega)+\chi)} e^{\sum_{i=1}^\ell \alpha_i\int_{o}^{z_i} \Omega} e^{-2\pi\ii(\epsilon(\Omega),\mathfrak a(D))}.
\eeq

\ed

\bl
If $\Omega$ has poles with integer residues, then the Szeg\"o kernel has no monodromies:
\beq
\psi([z_1+\gamma]-[z_2];\Omega) = \psi([z_1]-[z_2];\Omega) = \psi([z_1]-[z_2+\gamma];\Omega)
\eeq
for all cycle $\gamma$.

\el

\proof{
$\bullet$ If $\gamma=\sum_{i=1}^\genus n_i \acycle_i$ we have 
\beq
\mathfrak a(z_1+\gamma) = \mathfrak a(z_1)+ \mathbf n
\qquad \text{ where }\mathbf n=(n_1,\dots,n_\genus)
\eeq
and $\Theta(u+\mathbf n)=\Theta(u)$, this implies
\bea
\psi([z_1+\gamma]-[z_2];\Omega)
&=& \psi([z_1]-[z_2];\Omega) \ e^{\int_\gamma \Omega} \ e^{-2\pi\ii (\epsilon(\Omega),\mathbf n)} \cr
&=& \psi([z_1]-[z_2];\Omega),
\eea
because
\beq
\int_\gamma \Omega
= \sum_{i=1}^\genus n_i \oint_{\acycle_i} \Omega = 2\pi\ii \sum_{i=1}^\genus n_i \epsilon_i(\Omega) = 2\pi\ii (\epsilon(\Omega),\mathbf n).
\eeq

$\bullet$ If $\gamma=\sum_{i=1}^\genus n_i \bcycle_i$ we have 
\beq
\mathfrak a(z_1+\gamma) = \mathfrak a(z_1)+ \tau \mathbf n
\qquad \text{ where }\mathbf n=(n_1,\dots,n_\genus)
\eeq
and $\Theta(u+\tau \mathbf n)=\Theta(u) e^{-2\pi\ii ((u,\mathbf n) + \frac12 (\mathbf n,\tau \mathbf n) ) }$, this implies
\bea
\psi([z_1+\gamma]-[z_2];\Omega)
&=& \psi([z_1]-[z_2];\Omega) \ \frac{e^{-2\pi\ii ((\zeta(\Omega)+\mathfrak{a}(D)+\chi,\mathbf n) + \frac12 (\mathbf n,\tau \mathbf n) )}}{e^{-2\pi\ii ((\mathfrak{a}(D)+\chi,\mathbf n) + \frac12 (\mathbf n,\tau \mathbf n) )}} \ e^{\int_\gamma \Omega} \ e^{-2\pi\ii (\epsilon(\Omega),\tau \mathbf n)} \cr
&=& \psi([z_1]-[z_2];\Omega) \ e^{-2\pi\ii (\zeta(\Omega),\mathbf n)}  \ e^{\int_\gamma \Omega} \ e^{-2\pi\ii (\epsilon(\Omega),\tau \mathbf n)} \cr
&=& \psi([z_1]-[z_2];\Omega) 
\eea
because
\beq
\int_\gamma \Omega
= \sum_{i=1}^\genus n_i \oint_{\bcycle_i} \Omega = 2\pi\ii (\zeta(\Omega)+\tau \epsilon(\Omega),\mathbf n).
\eeq

$\bullet$ If $\gamma$ is a small circle around a pole $p$ of $\Omega$ with residue $\Res_p \Omega = t_{p,0}(\Omega)$ we have 
\bea
\psi([z_1+\gamma]-[z_2];\Omega)
&=& \psi([z_1]-[z_2];\Omega) \ e^{2\pi\ii t_{p,0}(\Omega)} \cr
&=& \psi([z_1]-[z_2];\Omega) 
\eea
because we assumed $t_{p,0}(\Omega)\in \mathbb Z$.
}

\section{1st construction: $\Gamma=$ space of meromorphic forms}
\label{sec:Tau1}

Here we present the so called ``Theta-function" solution of Fay identities, also called the finite-gap Tau function or Isospectral Tau function, or reconstruction formula.

Let $\curve$ a compact Riemann surface of genus $\genus$, equipped with a Jordan loops marking, and with a degree $d$ covering map $\x:\curve\to \curverond$.

The Abelian space $\Gamma$ considered in this 1st construction is the vector space of meromorphic 1-forms on $\curve$, modulo 1st kind and 3rd kind forms.

\bd

Let $\mathcal P$ a finite set of points of $\curve$, called the ``poles" (chosen out of the support of Jordan loops markings).
Let
\beq
\mathfrak M^1_{\mathcal P}(\curve) := \{ \Omega \in \mathfrak M^1(\curve) \ | \ \Omega \text{ has poles only at } \mathcal P \text{ or no pole}\}.
\eeq
\beq
\mathfrak M^1_{\mathcal P}(\curve)''' := \{ \Omega \in \mathfrak M^1(\curve) \ | \ \Omega \text{ has only simple poles at } \mathcal P \text{ or no pole}\}.
\eeq

\beq
\Gamma_{\mathcal P} := \mathfrak M^1_{\mathcal P}(\curve)/\mathfrak M^1_{\mathcal P}(\curve)'''.
\eeq
With the time map on $\Gamma_{\mathcal P}$:
\bea
\Omega & \mapsto & \{t_{p,k}(\Omega)\}_{p\in \mathcal P, \ k\geq 1}.
\eea
For any $\Omega$, only finitely many times are non-zero.
Times are coordinates on $\Gamma_{\mathcal P}$.

\ed

\bp[Preimage of the Sato vector]
Let $p\in \mathcal P$. Let $U_{p}\subset\curve$ an open simply connected neighborhood of $p$ in which the canonical local coordinate $\xi_{p}$ is defined.

Let $z\in \curve$.
If $z\in U_{p}$, let $\xi=\xi_{p}(z)$.

The 3rd kind differential of \autoref{lemdef:3rdkinddif},  $\omega'''_{z,p}$ is the preimage of the Sato vector $[\xi_p(z)]$. 
Indeed as a formal series of $\xi$, we have
\beq
\vec t_p(\omega'''_{z,p}) = [\xi] = [\xi_p(z)] .
\eeq
If $D=\sum_{i=1}^\ell \alpha_i.z_i$ is a divisor whose points are in $U_p$, we have, as a formal series of $\xi_p(z_i)s$ the time map of the 3rd kind differential $\omega'''_D$ is the Sato vector $[D]$:
\beq
\vec t_p\left(\omega'''_D \right)= \vec t_p\left(\sum_{i=1}^\ell \alpha_i \omega'''_{z_i,p}\right) = [D] .
\eeq

\ep

\proof{
Let $z\in U_p $, and consider the 1-form $\omega'''_{z,p}(\td z)$, it behaves as
\beq
\omega'''_{z,p}(\td z) \sim 
\frac{d\xi_p(\td z)}{\xi_p(\td z)-\xi_p(z)} - \frac{d\xi_p(\td z)}{\xi_p(\td z)}
+ \text{analytic} 
\eeq
and notice that the times will not depend on the analytic part.

If $z$ is close to $p$, so that $|\xi_p(z)|<|\xi_p(\td z)|$ we have the Taylor expansion in $\CC[[\xi_p(z)]]$:
\beq
\omega'''_{z,p}(\td z) \sim 
\frac{d\xi_p(\td z)}{\xi_p(\td z)-\xi_p(z)} - \frac{d\xi_p(\td z)}{\xi_p(\td z)} 
\sim \sum_{k=1}^\infty \frac{\xi_p(z)^{k}}{\xi_p(\td z)^{k+1}} \ d\xi_p(\td z) .
\eeq
This implies that, in the sense of formal series in $\CC[[\xi_p(z)]]$, we have
\beq
t_{p,k}(\omega'''_{z,p}) = \xi_p(z)^{k}
\eeq
which shows that this is indeed the Sato vector.
}

\br
It is important to notice that $\omega'''_{z,p}(\td z)$ is a well defined meromorphic 1-form defined on $\td z \in \curve$, whereas  the formal germ
$\sum_{k=1}^\infty \frac{\xi_p(z)^{k}}{\xi_p(\td z)^{k+1}} \ d\xi_p(\td z)$ makes sense only in a  neighborhood of $p$, it is the Taylor series expansion in the neighborhood.

In other words, $\Omega=\omega'''_{z,p} $ is defined globally on $\curve$, whereas the writting as a Sato vector is a Taylor series germ only within a neighborhood.

\er

\br
The time map is invertible, on the quotient by holomorphic and 3rd kind 1-forms. Somehow in the map, the times ``forget" the period and 3rd kind coordinates.

\er

Usually in integrable systems, one wants to find a Tau function that is a function (or formal series) of times.
We shall rather consider a Tau function that depends on a 1-form $\Omega\in \mathfrak M^1(\curve) $:

\bd[Hirota - Fay equations]
\label{def:HiotaFaycaseIsospectral}

A function 
\beq
\Tau : \mathfrak M^1(\curve) \to \CC
\eeq
is said to satisfy Fay identities, iff:
\bea\label{eq:FayId1}
 \forall \ z_1,z_2,z_3,z_4 \in \curve && \cr
\Tau(\Omega)\Tau(\Omega+\omega'''_{z_1,z_2}+\omega'''_{z_3,z_4}) 
&=& 
\Tau(\Omega+\omega'''_{z_1,z_2})\Tau(\Omega+\omega'''_{z_3,z_4}) \cr
&& - \Tau(\Omega+\omega'''_{z_1,z_4})\Tau(\Omega+\omega'''_{z_3,z_2}).
\eea

In other words, $\Tau$ is a section of a line bundle $\mathcal L \to \mathfrak M^1(\curve)$ that descends to a Tau function  on $\Gamma_{\mathcal P}$ in the times coordinates
\bea
& \mathcal L & \cr
& \downarrow &  \ \uparrow  \Tau = \text{section} \cr
& \mathfrak M^1(\curve) & \cr
& \downarrow & \mathbf t=\{\vec t_p\}_{p\in \mathcal P} \cr
& \Gamma_{\mathcal P} & 
\eea

\ed
Equation \eqref{eq:FayId1} has been solved by the so-called ``reconstruction formula" \cite{Krichever,Kri92,Kri92b,Kri77,Krichever_1978,Dub81}, using theta functions which we recall now.

\subsection{Theta Tau function}
\label{sec:TauTheta}

This 1st way of obtaining solutions to Hirota/Fay equations, is sometimes called ``finite gap solution" or ``isospectral Tau function", or the ``reconstruction formula".
It is mostly the work of the Russian school, particularly developed by Its, Matveev, Krichever, Novikov, Dubrovin, and many others \cite{Krichever,Kri77,itsMatveev75,RevRusInt} ...
We refer also to the textbook \cite{BBT}.

\bt[Fay's Theorem ``Theta functions satisfy Fay identities"]
\label{thm:FayTheta}
let $D=\sum_{i=1}^n [z_i]-[\td z_i] $ a supersymmetric divisor.
The Sezg\"o kernel satisfy the determinantal formula
\beq\label{eq:FayThetaId}
\psi(D;\Omega)
= \det_{i,j=1,\dots,n} 
\psi([z_i]-[\td z_j];\Omega).
\eeq
\et

\proof{Due to Fay \cite{Fay}. Sketching the idea of Fay's proof is that the ratio of both sides, viewed as a function of any point of the support of $D$, has no monodromy and no pole. Therefore it must be a constant, independent of the points of $D$. It suffices to evaluate it in a limit of coinciding points, and in this limit the ratio tends to 1. Notice that in genus zero eq~\eqref{eq:FayThetaId} reduces to the Cauchy determinant formula.}

As an immediate corollary we obtain the famous ``reconstruction formula" for Tau functions (see textbook \cite{BBT}):
\bt[Theta Tau function]
Let
\beq
\Tau(\Omega) = \Theta(\zeta(\Omega)+\chi) \ e^{\frac12 Q(\Omega,\Omega)} \ e^{-2\pi\ii (\epsilon(\Omega),\zeta(\Omega))} \ e^{-\pi\ii (\epsilon(\Omega),\tau\epsilon(\Omega))}
\eeq
where $Q$ is the quadratic form
\bea
Q(\Omega,\Omega) 
&=& \sum_{p\in \mathcal P} \sum_{k=1}^{-1+\deg_p\Omega} \frac{1}{k} t_{p,k}(\Omega) \Res_p \xi_p^{-k}\Omega \cr
&& +\sum_{p\in \mathcal P} t_{p,0}(\Omega) \int_o^p \Omega  
 +  \sum_{i=1}^\genus \epsilon_i(\Omega) \oint_{\bcycle_i} \Omega 
.
\eea
It is such that
\beq
\frac{\Tau(\Omega+\omega'''_D)}{\Tau(\Omega)} = \psi(D;\Omega).
\eeq

And, viewed as a function of the times $\mathbf t(\Omega)$, $\Tau(\Omega)$  satisfies the Fay determinantal formulas of \autoref{def:HiotaFaycaseIsospectral}, and therefore it satisfies Hirota equations for each family of times $\vec t_p(\Omega)$ associated to each $p\in\mathcal P$.

\et

\proof{
An easy computation yields
\beq
\zeta(\Omega+\omega'''_D) = \zeta(\Omega) + \mathfrak a(D).
\eeq
Let us define
\bea
\td Q(\Omega,\Omega) 
&=& Q(\Omega-2\pi\ii(\epsilon(\Omega),\omega'),\Omega-2\pi\ii(\epsilon(\Omega),\omega')) \cr
&=& Q(\Omega,\Omega) - 4\pi\ii (\epsilon(\Omega),\zeta(\Omega)) - 2\pi\ii (\epsilon(\Omega),\tau\epsilon(\Omega)),
\eea
this shows that
\beq
\Tau(\Omega) = \Theta(\zeta(\Omega)+\chi) \ e^{\td Q(\Omega,\Omega)}.
\eeq
We have
\beq
\td Q(\Omega+\omega'''_D,\Omega+\omega'''_D) = \td Q(\Omega,\Omega)
+ 2\sum_{i=1}^\ell \alpha_i \int_o^{z_i} \Omega   -2\ln{E(D)},
\eeq
this gives
\beq
\frac{\Tau(\Omega+\omega'''_D)}{\Tau(\Omega)} = \psi(D;\Omega).
\eeq

Then \autoref{thm:FayTheta}, implies that $\Tau$ is a Tau function solution to Fay identities and thus Hirota equations.
}

\section{2nd construction: Spectral curves}
\label{sec:Tau2}

We shall now present another possible solution of Fay identities.

Let the base curve $\curverond$ be a compact Riemann surface, we don't necessarily choose $\curverond=\CC P^1$.

\subsection{Spectral curves}

Here we take the following definition\footnote{alternative definitions or generalizations exist, in particular generalizations for covering maps $\x$ of infinite degree, or for non compact surfaces or non connected,...etc}:
\bd[Spectral curve]
A spectral curve is a meromorphic Lagrangian immersion of a compact Riemann surface $\curve$  of some genus $\genus$ into the total cotangent bundle $T^*\curverond$ of the base curve $\curverond$.
Let $\ii=(\x,\y)$ the immersion $\ii:\curve\hookrightarrow T^*\curverond$, where $\x$ is the projection to the base, and $\y$ the value in the fiber ($\y$ is thus the restriction of the tautological 1-form of $T^*\curverond$ to the image of the curve, $\y$ is called the Liouville form):
\bea
\curve & \overset{\ii}{\hookrightarrow} & T^*\curverond \cr
\x & \searrow & \downarrow \cr
& & \curverond
\eea
The base projection $\x:\curve\to\curverond$ is a holomorphic  ramified covering map.
$\y$ is a meromorphic 1-form  on $\curve$: 
\beq
\y\in\mathfrak M^1(\curve).
\eeq

Useful remark: when $\curverond=\CC P^1$, we can use the 1-form $d\x$ to identify the fiber of $T^*\CC P^1$ with $\CC$, i.e. write the 1-form $\y$ as 
\beq
\y = yd\x \qquad , \ \ \text{fiber of } T^*\CC P^1 \sim \CC.
\eeq
I.e. we can replace the meromorphic 1-form $\y$ by a meromorphic function $y$.
\ed

\bd[Times of a spectral curve]
The times of a spectral curve $\spcurve$, are defined as the times of the 1-form $\y=y d\x$.
Let $\mathcal P=\{ \text{poles of }\y\}$, these are called the punctures
\beq
\mathcal P = \{\text{punctures}\} = \{ \text{poles of }\y\}.
\eeq
\beq
\mathbf t = \mathbf t(\spcurve) = \{t_{p,k}\}_{p\in \mathcal P,  \ k\geq 1 } \ , \qquad t_{p,k} = t_{p,k}(\spcurve) = t_{p,k}(\y) = \Res_{p} \xi_p^k \y.
\eeq
Since $\y$ is meromorphic, there are only finitely many punctures, and finitely many non-vanishing times.
\ed

\bd[Periods of the spectral curve]
We define the ``periods" for all $j\in[1,\dots,\genus]$.
\beq
\epsilon_j =\epsilon_j(\spcurve)  =  \epsilon_j(\y) := \frac{1}{2\pi i} \oint_{\acycle_j} \y  \qquad \left( = \frac{1}{2\pi i} \oint_{\acycle_j} y d\x\right)
\eeq
where we wrote the last expression in the case $\curverond=\CC P^1$.

Periods associated to punctures are called \textbf{3rd kind periods} or \textbf{3rd kind times}, and are worth:
\beq
t_{p,0} = t_{p,0}(\spcurve) = t_{p,0}(\y) := \Res_{p} \y  \qquad \left( = \Res_p y d\x \right).
\eeq

\ed

As we shall see these periods of 1st and 3rd kind are associated to monodromic datas, they won't obey isomonodromic deformation equations.

\bp[Sato shift]
Let $D =\sum_{i=1}^{\ell} \alpha'_i.z_i $ a divisor of points of $\curve$ (whose support is assumed disjoint from the punctures, marked Jordan loops and ramification or nodal points).

There exists $r>0$ and there exists a 1-parameter family of spectral curves $\spcurve_{u}$ for $u\in [0,r[$, with immersions $(\x_u,\y_u):\curve_u\hookrightarrow T^*\curverond$, 
such that:

$\bullet$ $\forall u\in [0,r[$, the punctures $\mathcal P_u$ (poles of $\y_u$) are in holomorphic bijection with those of $\mathcal P$, with the same $\x$ projection and with same degrees, i.e. there is a continuous family of  bijections $p\mapsto p_u$ for each $u$, such that
\beq
\x_u(p_u) = \x(p),\quad
\deg_{p_u} \y_u = \deg_p \y,\quad
p_{u=0}=p.
\eeq

$\bullet$ $\forall u\in [0,r[$ there exists a continuous family of divisor $D_u = \sum_{i=1}^\ell \alpha'_i .\zeta_u(z_i)$ of $\curve_u$ such that
\beq
\forall i = 1,\dots, \ell \qquad \x_u(\zeta_u(z_i)) = \x(z_i),
\qquad  \ \ \zeta_{u=0}(z_i)=z_i.
\eeq

$\bullet$ $\curve_u\setminus (\mathcal P_u \cup \operatorname{supp} D_u)$ is isotopic to $\curve \setminus (\mathcal P \cup \operatorname{supp} D)$,

$\bullet$ and in fibers of $T^*\curverond$
\beq
\forall u\in [0,r[
\quad \frac{d}{d u} \y_u = \omega'''_{D_u}.
\eeq

Its times are worth
\beq
t_{p,k}(\spcurve_u) = t_{p,k}(\spcurve) + \delta_{k,0} \sum_{i=1}^\ell \alpha'_i (\delta_{p,\zeta_u(z_i)} -\delta_{p,\infty})
\eeq
and with constant periods
\beq
\epsilon_i(\spcurve_u) = \epsilon_i(\spcurve). 
\eeq

\smallskip

We denote it:
\beq
\spcurve_u = \spcurve+u[D].
\eeq
\ep

We admit this proposition here without proof, we refer to \cite{Eyn17, EynTgSp}. We give an illustrative example below in \autoref{sec:examplespcurvefamily}.

\bl[Associativity of the shift]
The Sato addition is associative and commutative.

If $D$ and $D'$ are two supersymmetric divisors, we have
\beq
\spcurve+u(D+D') = (\spcurve+ u D) +  u D'_u = (\spcurve+ u D') +  u D_u.
\eeq
If $D=\sum_{i=1}^\ell \alpha_i z_i$ we have
\beq
\vec t_p(\spcurve+ u D) = \vec t_p(\spcurve) + u \sum_{i=1}^\ell \alpha_i [\xi_p(z_i)].
\eeq
\el

Instead of giving the proof of the proposition and lemma, let us show how it works on an example:

\subsubsection{Prototypical example}
\label{sec:examplespcurvefamily}

Take $\curverond=\CC P^1$ and $\curve=\CC P^1$ and consider the spectral curve $\spcurve$ (called Airy Spectral curve) defined by the immersion
\beq
\spcurve :
 \left\{
\begin{array}{lll}
\x(z) & = & z^2 \cr
\y(z) & = & 2 z^2 dz
\end{array}
\right.  
\eeq
or in other words with $\y=y d\x$, and $d\x=2z dz$, we have
\beq
y(z) = z.
\eeq
It satisfies the algebraic equation:
\beq
y^2-x=0.
\eeq

There is a unique puncture $\mathcal P=\{\infty\}$, and its only non-vanishing time is:
\beq
t_{\infty,3}(\spcurve) = -2.
\eeq

Let $z_1\in \curve$, and $X_1 = \x(z_1)=z_1^2 \in \curverond$.
Let us define the function $\zeta_u(z_1)$ as the solution of
\beq
{\zeta_u}^2 + \frac{u}{\zeta_u} = z_1^2
\eeq
and which behaves as $\zeta_u(z_1) = z_1 + O(u)$ at small $u$.
The solution as a series is
\beq
\zeta_u(z_1) = z_1 - \frac12 \sum_{k=1}^\infty \frac{(\frac32(k-1))!}{k! \ (\frac12(k-1))!} \ z_1^{1-3k}\ u^k.
\eeq
Its radius of convergence is
\beq
|u| < \frac{2 |z_1|^3 }{3\sqrt{3}}.
\eeq

The spectral curve
$\spcurve_u = \spcurve + u [z_1] = (\x_u,\y_u)$
is defined by the immersion
\beq
\spcurve +u [z_1] :
 \left\{
\begin{array}{l}
\x_u(z) = z^2 + \frac{u}{\zeta_u(z_1)} \cr
\y_u(z) = \left( z + \frac{u}{2\zeta_u(z_1)(z-\zeta_u(z_1))} \right) 2z dz
\end{array}
\right.  .
\eeq
It satisfies the equation
\beq
(y^2-x)(x-X_1)= u \left(y +\zeta_u(z_1)-\frac{u}{4\zeta_u^2(z_1)}\right) .
\eeq
It indeed reduces to $y^2-x=0$ when $u=0$.

It has the times
\beq
t_{\infty,k}(\spcurve_u) = -2 \delta_{k,3} = t_{\infty,k}(\spcurve)
\qquad \forall \ k>0
\eeq
and
\beq
t_{\zeta_u(z_1),0}(\spcurve_u)=-t_{\infty,0}(\spcurve_u) = u.
\eeq

More generally, if $D=\sum_{i=1}^\ell \alpha'_i [z_i]$, the spectral curve $\spcurve_u = \spcurve+u D$ is given by:
\beq
\spcurve_u=\spcurve +u D :
 \left\{
\begin{array}{l}
\x_u(z) = z^2 + u C_u \cr
\y_u(z) = \left( z + \frac{u}2 \sum_{i=1}^\ell \frac{\alpha'_i}{\zeta_u(z_i)(z-\zeta_u(z_i))} \right) 2z dz
\end{array}
\right. 
\eeq
where $C_u$ and $\zeta_u(z_i)$ for $i=1,\dots,\ell$, are defined by the $\ell+1$ equations

\bea
\zeta_u(z_i)^2 &=& z_i^2- u C_u , \ \ \zeta_u(z_i) = z_i +O(u) \cr
C_u &=& \sum_{i=1}^\ell \frac{\alpha'_i}{\zeta_u(z_i)}.
\eea
In fact it is easy to see that $\spcurve+uD$ reduces to $\spcurve$ at $u=0$, and has the times
\beq
t_{p,k}(\spcurve+uD) = t_{p,k}(\spcurve) + u \delta_{k,0} \sum_{i=1}^\ell  \alpha'_i (\delta_{p,\zeta_u(z_i)}-\delta_{p,\infty}).
\eeq

\subsubsection{Spectral curve Tau function}

\bd[Spectral curve Tau function]
We want to find a function defined on the ``moduli-space of spectral curves" (we don't define this moduli space here, see \cite{Eyn17,EynCycles,EynTgSp}), i.e. a section of a line bundle over this moduli space 
\beq
\Tau(\spcurve)
\eeq
that satisfies Fay identity

\beq\label{eq:FayeqSpCurve}
 \forall \ D=\sum_{i=1}^n [z_i]-[\td z_i] \text{ supersymmetric} \quad 
\frac{\Tau(\spcurve+D)}{\Tau(\spcurve)} 
= 
\det \frac{\Tau(\spcurve+[z_j]-[\td z_i])}{\Tau(\spcurve)} .
\eeq

\ed

The solution to these Fay identities can in general not be written with classical functions (like Theta functions), they are genuinely transcendental, having the Painlevé property.

However, they can sometimes be expressed as integrals (matrix integrals), and in general they can be expressed as power series in some large times limit (also called heavy charge limit), as formal series expansion whose terms are computed by Topological Recursion, that we show in \autoref{sec:TauTR} below.

\subsection{Spectral curves Tau function}
\label{sec:TauTR}

We shall describe a solution to \eqref{eq:FayeqSpCurve} as an asymptotic series, thanks to Topological Recursion \cite{EO07,Eynard:2014zxa}.

Topological Recursion was introduced initially to compute large order asymptotic expansions of matrix integrals \cite{Eynard:2002kg,Eynard_2004,Chekhov:2005rr,Chekhov:2006vd}, and was then turned into a recursive definition of algebraic invariants \cite{EO07,Borot_2020},

\smallskip

Let $\hbar$ a ``small" parameter, it is also sometimes called the ``quantum" parameter, or the ``dispersion" parameter, that will scale the spectral curve. The limit $\hbar$ small will correspond to ``large spectral curves" or ``heavy spectral curves".
\bd[Rescaling]
We define the rescaling by $\lambda\in \CC^*$ of a spectral curve $\spcurve=\curve\hookrightarrow T^*\curverond$ with immersion map $(\x,\y)$, as the rescaling of the fiber coordinate $\y\to \lambda \y$:
\beq
\lambda\spcurve:=(\curve,(\x,\lambda\y)).
\eeq
This rescales all times and periods:
\beq
t_{p,k}(\lambda\spcurve) = \lambda \ t_{p,k}(\spcurve), \quad
\epsilon_i(\lambda\spcurve) = \lambda \ \epsilon_i(\spcurve).
\eeq
\ed
In particular, the spectral curve $\hbar^{-1}\spcurve$ has ``large times" in the limit $\hbar\to 0$, it is called a ``\textbf{heavy  limit}".

The conjecture presented in \cite{EO07, BoEy10, Eyn17,E08,EM08} is that topological recursion yields a solution of Fay/Hirota \eqref{eq:FayeqSpCurve}, as formal series in $\hbar$:

\begin{conjecture}[Tau functions from topological recursion]
Let
\beq
\hat\Theta(\vec u;\tau) = e^{\hbar^{-2}F_0(\spcurve)+F_1(\spcurve)}  e^{-2\pi\ii (\hbar^{-1} \epsilon-\mu,\vec u)} 
 \Theta(\vec u;\tau)
e^{-\pi\ii \hbar^{-2} (\epsilon,\tau\epsilon)}
 e^{2\pi\ii (\hbar^{-1}\epsilon,\chi)}
  e^{-2\pi\ii (\mu,\chi)} e^{\pi \ii (\mu,\tau\mu)}
\eeq
where $\chi=\nu+\tau \mu$.

In \cite{BoEy10,Eyn17,E08,EM08} is defined the following formal Tau function:
\beq
\label{eq:defTauhbarSpcurve}
\Tau(\hbar^{-1}\spcurve)
= \Big( 1 + \sum_{k=1}^\infty \frac{1}{k!}
\sum_{\overset{\vec m_i, \ g_i}{2g_i+|\vec m_i|-2>0}} \prod_{i=1}^k \frac{\hbar^{2g_i+|\vec m_i|-2}}{\prod_{j=1}^\genus m_{i,j}!} F_{g_i}^{(\vec m_i)}(\spcurve) \partial^{\vec m_i}_{\vec u} \Big) \left. \hat\Theta(\vec u;\tau) \right|_{\vec u=\hbar^{-1}\zeta(\spcurve)+\chi}
\eeq
where $F_g(\spcurve)$ and $\omega_{g,n}(\spcurve)$ are the TR invariants of $\spcurve$ (see \cite{EO07, Eyn17}),
and  where $\vec m_i=(m_{i,1},\dots,m_{i,\genus})\in \ZZ_+^\genus $ and $|\vec m_i| = \sum_{j=1}^\genus m_{i,j}$, and
\beq
F_{g}^{(\vec m)}(\spcurve)
= \overbrace{\oint_{\bcycle_1}\dots\oint_{\bcycle_1}}^{m_1}
\overbrace{\oint_{\bcycle_2}\dots\oint_{\bcycle_2}}^{m_2}
\dots
\overbrace{\oint_{\bcycle_\genus}\dots\oint_{\bcycle_\genus}}^{m_\genus} 
\omega_{g,|\vec m|}(\spcurve),
\eeq
\beq
\partial^{\vec m}_{\vec u} = \prod_{j=1}^\genus \frac{\partial^{m_j}}{\partial u_j^{m_j}}.
\eeq

$\Tau(\hbar^{-1}\spcurve)$ is a formal series of $\hbar$ times an exponential $e^{\hbar^{-2}F_0(\spcurve)}$, and whose coefficients are differential polynomials of $\hat\Theta$. We choose a grading in $\hbar$ for which $\hat\Theta$ and its derivatives have degree 0.
For each power of $\hbar$ there is a finite number of terms.

\medskip

The 1st few terms are:
\bea
\Tau(\hbar^{-1}\spcurve)
& = & \hat\Theta(\hbar^{-1}\zeta(\spcurve)+\chi) \cr
&& + \hbar \Big( \sum_{i=1}^\genus \left(\oint_{\bcycle_i} \omega_{1,1}(\spcurve)\right) {\hat\Theta}'^{(i)} + \frac{1}{6}\sum_{i,j,k=1}^\genus \left(\oint_{\bcycle_i}\oint_{\bcycle_j}\oint_{\bcycle_k} \omega_{0,3}(\spcurve)\right) {\hat\Theta}'''^{(i,j,k)}  \Big) \cr
&& + O(\hbar^2).
\eea

The conjecture is that this $\Tau$ satisfies Fay identities \eqref{eq:FayeqSpCurve}, order by order in $\hbar$.

\end{conjecture}

\textbf{Proof attempts:}

\begin{itemize}

\item In \cite{BoEy10} it was verified that the conjecture holds true for the first 3 orders in $\hbar$, for arbitrary compact spectral curves.

\item Using Hirota/Fay vs Lax pairs.
There has been proofs to all orders in $\hbar$ for certain families of spectral curves, using a Lax pair differential system, also related to the notion of quantum curve.

Two kinds of proofs exist in the literature: either start from Topological Recursion (TR) and derive Fay identities order by order as formal series, or vice-versa, start from a Tau function satisfying Fay, and proving that its formal asymptotic expansion obeys TR.

Cases where it is proved:

\begin{itemize}

\item Genus zero spectral curves are easier (there is no Theta-function term):

\begin{itemize}

\item Airy curve: Berg\`ere, Eynard 2009:  ``Determinantal formulae and loop equations" \cite{BE09}.
In this article a version of the Fay identity is proved for the case $n=2$.

\item Genus zero ``topological type": Berg\`ere, Borot, Eynard 2013: ``Rational Differential Systems, Loop Equations, and Application to the $q$th Reductions of KP" \cite{Bergere:2013qba}.

\item Genus zero for ribbon graphs:  Do, Manescu, 2013:     ``{Quantum curves for the enumeration of ribbon graphs and hypermaps}" \cite{DoMa14}.

\item Genus zero Schr\"odinger: 
 Mulase, Sulkowski 2015:
``{Spectral curves and the Schr\"odinger equations for the Eynard-Orantin recursion}"
\cite{MS15}.

\item Review Norbury 2016: ``Quantum curves and topological recursion" \cite{N16}.

\item Genus zero   ``topological type": Belliard, Eynard, Marchal 2016: ``Integrable differential systems of topological type and reconstruction by the topological recursion" \cite{Belliard:2016cyc}, using ``{Loop Equations from Differential Systems on Curves}" \cite{Belliard:2016mxj}.

\item Genus zero for ``admissible spectral curves": Bouchard, Eynard 2016: ``Reconstructing WKB from topological recursion"  \cite{BouchEynWKB}. In this article a version of the Fay identity is proved for the case $n=2$.

\item Genus zero Hurwitz-type: Do, Dyer, Mathews 2017, ``Topological recursion and a quantum curve for monotone Hurwitz numbers" \cite{NDD17}.

\item
Dunin-Barkowski, Mulase, Norbury, Popolitov, Shadrin 2017:
``{Quantum spectral curve for the Gromov--Witten theory of the complex projective line}"
\cite{dunin2017quantum}.

\item Genus zero hypergeometric type: Bychkov, Dunin-Barkowski, Kazarian, Shadrin 2021 ``Topological recursion for Kadomtsev-Petviashvili tau functions of hypergeometric type" \cite{bychkov2021topological}. In this article the authors relate TR to KP, which satisfies Hirota-Fay identities.

\item Genus 0:
Kidwai, Osuga 2022 : ``{Quantum curves from refined topological recursion: the genus 0 case}" \cite{kidwai2022quantum} .

\item Weighted-Hurwitz numbers: Alexandrov, Chapuy, Eynard, Harnad, 2018 "Weighted Hurwitz numbers and topological recursion" \cite{Alexandrov:2018ncq,Alexandrov:2017nfs,Alexandrov:2016edh}.
In this series of articles the authors start from the Fay-Hirota property and derive the Toplogical Recursion as well as the quantum curve.

\end{itemize}

\item Higher genus:

\begin{itemize}

\item Higher genus: Dumitrescu, Mulase 2014:
``Quantum curves for Hitchin fibrations and the Eynard-Orantin theory", 
``Quantization of spectral curves for meromorphic Higgs bundles through topological recursion",   \cite{DM14a, DM14b}.

\item Genus 1: Iwaki, Marchal, Saenz 2016, 2017, 2018, 2020:
 ``{Quantum Curve and the First Painleve Equation}" \cite{IS16},
``{Painlev\'e equations, topological type property and reconstruction by the topological recursion}" \cite{iwaki2017painleve},
``{Painlevé equations, topological type property and reconstruction by the topological recursion}" \cite{I18},
 ``2-parameter $\tau$-function for the first Painlev{\'e} equation: topological recursion and direct monodromy problem via exact WKB analysis" \cite{iwaki20202}.

\item Hyperelliptic curves: Marchal, Orantin 2022:
``{Quantization of hyper-elliptic curves from isomonodromic systems and topological recursion}"
\cite{marchal2021quantization}, and Eynard Garcia-Failde 2023: ``{From topological recursion to wave functions and PDEs quantizing hyperelliptic curves}"
\cite{eynard_garcia-failde_2023}.
In this article a version of the Fay identity is proved for the case $n=2$.

\item Algebraic curves with only simple branch-points: 
Eynard, Garcia-Failde, Marchal, Orantin 2021: 
``{Quantization of classical spectral curves via topological recursion}"
\cite{eynard2021quantization}.
In this article a version of the Fay identity is proved for the case $n=2$.

\end{itemize}

\item Beyond:

\begin{itemize}
\item Difference equations: Marchal 2017: ``{WKB solutions of difference equations and reconstruction by the topological recursion}" \cite{Marchal_2017}.

\end{itemize}

\end{itemize}

\end{itemize}

And we might have forgotten some other proofs, in particular in cases of specific applications...

\section{Other Tau functions}
\label{sec:TauMm}
We shall not be exhaustive and mention only the following example: the 1--Matrix model prototype, which is extremely useful and used with many applications from combinatorics to quantum gravity and string theory.

\subsection{Matrix integrals}
\label{sec:matrices}

Typical examples of Tau functions are matrix integrals.
It is usual to consider the punctures at $\infty$ and write $x=1/\xi$, i.e. $\xi=0$ corresponds to $x=\infty$.

\medskip

Let $\Gamma = \CC[x]$ the vector space space of complex polynomials.
To $\vec t\in \Gamma $ associate the polynomial, called the ``potential"
\beq
V(x) = \sum_{k=1}^{\deg V} \frac{t_k}{k} x^k.
\eeq
Let $N$ a positive integer and $H_N$ the space of $N\times N$ Hermitian matrices, equipped with the Lebesgue measure $dM$.
Let:
\beq
\Tau_N(\vec t) = \int_{H_N} dM e^{-\Tr V(M)}
\eeq
where we assume that the integral is absolutely convergent (if it is not absolutely convergent on $H_N$, replace $H_N$ by $H_N(\gamma)$ the space of $N\times N$ normal matrices with eigenvalues on a Jordan arc $\gamma$, see \cite{BookRMT}).

For a divisor $D=\sum_i \alpha_i.z_i $, the Sato shift $\vec t\mapsto \vec t+[D]$ amounts to
\beq
V(x) \mapsto V(x) + \sum_{i} \alpha_i \ln (x-z_i),
\eeq
(notice that the Sato shifted potential is not a polynomial, it is a polynomial only order by order as formal series),
in other words
\beq
\Tau_N(\vec t+[D]) = \int_{H_N} dM e^{-\Tr V(M)} \ \prod_{i=1}^\ell \det(M-z_i)^{-\alpha_i}.
\eeq
In particular with the divisor $D=[\infty]-[x]$ we consider
\beq
\psi_N(\vec t;x) =  \frac{\Tau_N(\vec t-[x]+[\infty])}{\Tau_N(\vec t)} = \frac{1}{\Tau_N(\vec t)}\int_{H_N} dM e^{-\Tr V(M)} \ \det(x-M),
\eeq
and with the divisor $D=[x]-[\infty]$ we consider
\beq
\phi_N(\vec t;x) =  \frac{\Tau_N(\vec t+[x]-[\infty])}{\Tau_N(\vec t)} = \frac{1}{\Tau_N(\vec t)}\int_{H_N} dM e^{-\Tr V(M)} \ \frac{1}{\det(x-M)}.
\eeq
It is well known in the theory of random matrices \cite{kostov} that they satisfy an orthogonality relationship
\beq
\int_\RR \psi_n(\vec t;x) \phi_m(\vec t;x) dx = \delta_{n,m}.
\eeq
Moreover, it is well known \cite{Akemann04,Bergere03,Bergere04,Bergere05} that for supersymmetric divisors they satisfy Hirota equations
\beq
\frac{\Tau_N(\vec t+\sum_{i=1}^\ell [z_i]-[\td z_i])}{\Tau_N(\vec t)} = \det \frac{\Tau_N(\vec t+[z_i]-[\td z_j])}{\Tau_N(\vec t)}.
\eeq
Matrix integrals do satisfy Hirota equations and Fay identities.

More generally, matrix integrals are known to be Tau functions \cite{Bertola:2004ws,Bertola:2002aw,Harnad,BookRMT}.

\section{Conclusion}
\label{sec:conc}
This article presented a more modern and TR compatible reformulation of Hirota and Fay equations. 

- The usual formulation of Hirota equations is with formal series, and if we multiply by the appropriate exponential factor and differential form factor, this is formulated with a spinor trans-monomial.

- In the language of formal series or trans-monomials, Hirota equations are equivalent to Fay identities.
But then Fay identities can be formulated not only as formal series, they can be extended to functions on Riemann surfaces.

- Fay identities are then naturally formulated with Riemann surfaces, and some solutions can be found in terms of Riemann surfaces.

- We recalled two solutions, one where $\Tau$ is defined on the space of meromorphic forms on a fixed Riemann surface, and one where $\Tau$ is defined on the moduli space of spectral curves, and is given asymptotically by topological recursion.

- In Topological Recursion, people often mention the ``quantum curve conjecture" claiming that Tau functions Sato--shifted by unitary divisors should obey some ODE.
The ODE is in fact a consequence of integrability and thus the ``quantum curve conjecture" is in fact a consequence of a stronger ``integrability" conjecture, i.e. the fact that the Topological-Recursion Tau function eq~\eqref{eq:defTauhbarSpcurve} obeys Fay identities. This conjecture has been proved in number of cases but it is still to be proved for more general cases.

\section*{Acknowledgments}

This article is a result of the ERC-SyG project, Recursive and Exact New Quantum Theory (ReNewQuantum) which received funding from the European Research Council (ERC) under the European Union's Horizon 2020 research and innovation programme under grant agreement No 810573.
We thank R. Belliard for discussions, and O. Babelon and J. Harnad from whom I have learnt so many things about integrable systems.
I also want to thank M. Mulase, because this article reviews a way of thinking of Tau functions that emerged from a discussion with him in the Shinkansen between Tokyo and Kyoto in 2008, in the country of Hirota and Sato.

\printbibliography
\end{document}